\documentclass[twocolumn,showpacs,preprintnumbers,amsmath,amssymb,letterpaper]{revtex4}
\usepackage{amssymb}
\usepackage{graphicx}
\usepackage{dcolumn}
\usepackage{bm}

\begin{document}

\title{Master equation approach to the theory of diffusion in
alloys and calculations of enhancement factors for tracer solvent
and tracer solute diffusion in FCC alloys}
\renewcommand{\abstractname}{}
\author{ V. G. Vaks$^{a,b}$, A. Yu. Stroev$^{a,b}$,  I. R. Pankratov$^{a}$,
and I. A. Zhuravlev$^{a}$}

\date{\today}
\affiliation{$^{a}$National Research Center ``Kurchatov
Institute'', 123182 Moscow, Russia \\
$^{b}$Moscow Institute of Physics and Technology (State University),
117303 Moscow, Russia}

\begin{abstract}
\noindent{\bf Abstract} The earlier-suggested master equation
approach is used to develop the consistent statistical theory of
diffusion in substitution alloys  using the five-frequency model of
FCC alloys as an example. Expressions for the Onsager coefficients
in terms of microscopic interatomic interactions and some
statistical averages are presented. We discuss methods of
calculations of these averages using various statistical
approximations and both the nearest-neighbor and the second-shell
approximations to describe the vacancy correlation effects. The
methods developed are used for calculations of enhancement factors
for tracer solvent and tracer solute diffusion in dilute FCC alloys.
We show that some significant contribution to the tracer solvent
enhancement factor related to the thermodynamic activity of
vacancies was  missed in the previous treatments of this problem. It
implies that the most of existing estimates of parameters of
five-frequency model for real alloys should be revised. For the
tracer solute diffusion, the enhancement factor seems to be
calculated for the first time. The results obtained are used to
estimate the microscopic parameters important for diffusion,
including the vacancy-solute interaction, in several FCC alloys for
which necessary experimental data are available.
\end{abstract}


\maketitle

\section{INTRODUCTION\label{Introduction}}

Microscopic theories of diffusion in alloys can be divided into two
main groups: (i) those based on the random walk theory and the
``vacancy-solute association-dissociation'' models (to be called
``traditional'' theories)
\cite{Lidiard-55,Lidiard-60,Manning-64,Howard-Man-67,Bocquet-74,LeClaire-78,Ishioka-84,Allnatt-93,
Bocquet-96} and (ii) those based on the master equation approach
recently suggested by Nastar et al.
\cite{Nastar-00,Nastar-05,Barbe-06}. The traditional approaches
described in a number of reviews and textbooks
\cite{LeClaire-78,Allnatt-93,Bocquet-96} put grounds for the present
microscopic understanding of diffusion. However, the results of
these approaches are exact only in the dilute alloy limit when the
site fractions of solute, $c_{\alpha}$, tend to zero, while
extensions of these approaches to the finite $c_{\alpha}$ meet
difficulties. Even for a dilute binary alloy $AB$ with low $c_B\ll
1$, calculations of enhancement factors, that is, the linear in
$c_B$ terms in diffusion coefficients, for both the tracer solute
and the chemical (intrinsic) diffusion seem to be not performed,
while traditional calculations of the tracer solvent enhancement
factor, as discussed by Nastar \cite{Nastar-05} and below, include
significant inaccuracies.

The master equation approach (MEA) to the steady-state diffusion
theory suggested by Nastar et al.
\cite{Nastar-00,Nastar-05,Barbe-06}  can be used for both the dilute
and the concentrated  alloys. It can be formulated in terms of fully
microscopic notions, such as the ``saddle-point'', ``kinetic'' and
``configurational''  interactions discussed below which can be
calculated by $ab \ initio$ methods, see, e. g., \cite{SF-07}.
Nastar et al. \cite{Nastar-00} used MEA to study some general
features of diffusion in concentrated alloys, and the results
(obtained using rather simple approximations) reasonably agree with
Monte Carlo simulations. Nastar \cite{Nastar-05} used this approach
to study the long-discussed problem of calculating the tracer
self-diffusion enhancement factor, and her results agree with the
Monte Carlo simulations better than the traditional ones
\cite{Lidiard-60,Howard-Man-67,Ishioka-84}. Barbe and Nastar
\cite{Barbe-06} generalized the ``pairwise effective interaction''
version of MEA used in \cite{Nastar-00,Nastar-05} to study features
of diffusion in alloys with a high ratio of solute-vacancy to
solvent-vacancy exchange frequencies.

At the same time, the formulation of MEA given in Refs.
\cite{Nastar-00,Nastar-05,Barbe-06} includes a number of
shortcomings which   prevent from further applications of this
approach. First, even the most detailed paper \cite{Nastar-05} does
not contain explicit expressions for the basic quantity of the
theory, the vacancy-atom exchange probability, which hinders
understanding and generalizations of the method. Second, basic
equations in \cite{Nastar-05} are implicit and cumbersome, and it is
difficult to use them. Third and most important, treatment of
vacancy-solute interactions in Ref. \cite{Nastar-05} includes
inaccuracies which are manifested, in particular, in the
disagreement of one of results in \cite{Nastar-05} with that of
traditional theories, as discussed below in Sec. \ref{dilute
-binary}.

In this work we present a new formulation of the master equation
approach free from the shortcomings mentioned, and use it to
calculate the enhancement factors for tracer diffusion in dilute FCC
alloys. Our equations are explicit and simple, they can be solved
using the standard methods of statistical physics, and their
possible generalizations (for example, to the case of not-nearest or
non-pairwise interactions) are evident.

To be definite, we illustrate our approach by consideration of
dilute FCC alloys using the pairwise nearest-neighbor interaction
model which is commonly called ``the five-frequency model''
\cite{Lidiard-55,Lidiard-60,Manning-64,Howard-Man-67,Bocquet-74,LeClaire-78,Ishioka-84,Allnatt-93,
Bocquet-96}.  At the same time,  we also take into account the
solute-solute interactions not considered in the standard
five-frequency model.

The important general feature of our approach is the proper
description of effects of the vacancy-solute interaction  (or
``vacancy-solute binding energy''
\cite{Lidiard-55,Lidiard-60,Manning-64,Howard-Man-67,Bocquet-74,LeClaire-78,Ishioka-84,Allnatt-93,
Bocquet-96}), in particular, for the tracer self-diffusion
enhancement factor $b_{A^*}$. We show that all previous calculations
of $b_{A^*}$ using both the traditional methods
\cite{Lidiard-60,Howard-Man-67,Ishioka-84} and the version of MEA
used in \cite{Nastar-05}, missed the contribution of thermodynamic
activity of vacancies related to the vacancy-solute interaction.
This led to spreading of a pessimistic opinion that the ``diffusion
experiments by themselves are not sufficient to determine this
binding energy'' \cite{Bocquet-96}, and presently experimental
estimates of  this energy use various plausible models with no
consistent statistical justification
\cite{Faupel-88,Hagenschulte-89, Hagenschulte-94}. Our results show
that this pessimistic opinion is wrong, and the consistent estimates
of $v^{vB}$ for several alloys using available experimental data are
presented in this work. These results also imply that the most of
existing estimates of parameters of five-frequency model
(``frequency ratios'') for real alloys \cite{Allnatt-93,Bocquet-96}
should be revised.

The paper is organized as follows. In Sec. \ref{Gen-eqs} we present
main equations of the master equation approach. In Sec.
\ref{Gen-eq-for-Onsager}, these equatins are used to derive general
expressions for Onsager coefficients which describe the steady-state
diffusion in a substitution alloy. Here we use the methods suggested
by Nastar et al. \cite{Nastar-00,Nastar-05,Barbe-06} but employ our
explicit formulation of the master equation approach. In Sec.
\ref{Calc-averages-gen} we discuss  both exact relations and methods
of approximate calculations for statistical averages which enter
into the general expressions for Onsager coefficients. In Sec.
\ref{Diff-binary-gen} we present explicit expressions for  Onsager
and diffusion coefficients in a binary alloy at any concentration
and show that in the case of a dilute alloy, these expressions
coincide with those of the traditional theory \cite{Allnatt-93}. In
Sec. \ref{b_A^*-sec} and \ref{b_B^*-sec} we discuss the enhancement
factor for  diffusion of tracer solvent and tracer solute,
respectively. In Sec. \ref{estimates-of-x_n} we estimate parameters
of the five-frequency model and interactions significant for
diffusion for several alloys for which necessary experimental data
are available. We also find that the description of these data by
the five-frequency model seems to be physically reasonable. Our main
conclusions are summarized in Sec. \ref{Conclusions}.

\section{GENERAL EQUATIONS OF DIFFUSIONAL KINETICS IN SUBSTITUTION ALLOYS\label{Gen-eqs}}

Basic equations of the master equation approach for the diffusional
kinetics of substitution alloys have been derived earlier
\cite{BV-98,VZh-12}. Below we present the necessary relations from
Ref. \cite{VZh-12}. We consider a substitution alloy with $(m+1)$
components \,$p'$\, which include host atoms denoted by index $h$,
solute atoms denoted by Greek letters $\alpha$, $\beta$, $\lambda$,
$\mu$, $\nu$, and vacancies denoted by $v$. Latin letters $p$, $q$,
$r$ will denote all kinds of atoms, both $h$ and $\alpha$, while
Greek letters $\rho$, $\sigma$, $\tau$ will denote both solute atoms
$\alpha$ and vacancies $v$, thus the whole set $p'$ can be written
either as $\{p,v\}$ or as $\{h,\rho\}$\,. Distributions of atoms
over lattice sites $i$ are described by the different occupation
number sets $\{n_i^{p'}\}$ where the operator $n_i^{p'}$ is 1 when
the site $i$ is occupied by a \,$p'$-species component, and 0
otherwise. At each \,$i$\, these operators  obey the identity
\,$\sum_{p'}n_i^{p'}=1$.\, Hence only \,$m$\, of them are
independent, and one of these operators can be expressed via other
ones. We eliminate the host atom occupation operator $n_i^h$ writing
it as
\begin{equation}
n_i^h=\Big(1-\sum_{\rho}n_i^{\rho}\Big).\label{n^h_i}
\end{equation}
This is convenient to describe real alloys where the vacancy site
fraction is very low: $\langle n^v_i\rangle\ll\langle
n_i^{\alpha}\rangle$, while Nastar et al.
\cite{Nastar-00,Nastar-05,Barbe-06} eliminate vacancy occupation
operators $n_i^v$.

We use the pairwise interaction model for which the total
configurational Hamiltonian $H^t$ can be expressed via $n_i^{p'}$
and couplings $V_{ij}^{p'q'}$ as follows:
\begin{equation}
H^t=\sum_{ij}\Big({1\over 2}\sum_{pq}V_{ij}^{pq}n_i^{p}n_j^{q}+
\sum_{p}V_{ij}^{pv}n_i^{p}n_j^{v}+{1\over
2}V^{vv}_{ij}n_i^{v}n_j^{v}\Big).\label{H^t}
\end{equation}
After elimination of operators $n_i^h$ according to Eq
(\ref{n^h_i}), the Hamiltonian $H^t$ takes the form:
\begin{equation}
H^t= E_0+\sum_{\rho i}\varphi_{\rho} n_i^{\rho}
+H_{int}.\label{H^t-H_int}
\end{equation}
Here constants $E_0$ and $\varphi_{\rho}$ yield some insignificant
shifts in the total energy and chemical potentials while the
interaction Hamiltonian $H_{int}$ can be written as
\begin{equation}
H_{int}=\sum_{\alpha\beta,i>j}v_{ij}^{\alpha\beta}n_i^{\alpha}n_j^{\beta}+
\sum_{\alpha,ij}v_{ij}^{\alpha v}n_i^{\alpha}n_j^{v} \label{H_int}
\end{equation}
where terms $v_{ij}^{vv}n_i^{v}n_j^{v}$ with vacancy-vacancy
interactions are neglected, and the configurational interaction
$v_{ij}^{\alpha\rho}$ is expressed via couplings $V_{ij}^{p'q'}$ in
(\ref{H^t}) as follows:
\begin{equation}
v_{ij}^{\alpha\rho}=(V^{\alpha\rho}-V^{\alpha{h}}-V^{{h}\rho}
+V^{hh})_{ij}\,.\label{v_ij-def}
\end{equation}

The fundamental master equation for the probability $P$ of finding
an occupation number set $\{n_i^{\rho}\}=\xi$ can be written as
\cite{BV-98}:
\begin{equation}
dP(\xi)/ dt=\sum_{\eta} [W(\xi,\eta)P(\eta)-
W(\eta,\xi)P(\xi)]\equiv\hat SP\label{dP/dt}
\end{equation}
where \,$W(\xi,\eta)$\, is the $\eta\rightarrow\xi$ transition
probability per unit time. Adopting for probabilities \,$W$\, the
conventional ``transition state'' model \cite{BV-98,SF-07}, we
express the transfer matrix \,$\hat S$\, in  (\ref{dP/dt}) in terms
of the probability of an elementary inter-site atomic exchange
(``jump'') p$i\rightleftharpoons {v}j$\, between neighboring sites
\,$i$\, and \,$j$:\,
\begin{equation}
W_{ij}^{ pv}=n_i^{p}n_j^{v}\omega_{pv}^{eff}\exp[-\beta(\hat
E_{{p}i,{v}j}^{SP}-\hat E_{{p}i,{v}j}^{in})].\label{W_ij^pv}
\end{equation}
Here   \,$\beta =1/T$\, is the reciprocal temperature, \,$\hat
E_{{p}i,{v}j}^{SP}$\, is the saddle point energy, $\hat
E_{{p}i,{v}j}^{in}$ is the initial (before the jump) configurational
energy of a jumping atom and a vacancy, and  the factor
\,$\omega_{pv}^{eff}$\, can be written as
\begin{equation}
\omega_{pv}^{eff}=\omega_{pv}\exp\,\big(\Delta
S_{{p}i,{v}j}^{SP}\big)\label{omega_pv}
\end{equation}
where \,$\omega_{pv}$\, is the attempt frequency (which has the
order of magnitude of a mean frequency of vibrations of a jumping
atom in an alloy), and  \,$\Delta S_{{p}i,{v}j}^{SP}$\, is the
entropy difference between the saddle-point and initial alloy
states.

The saddle point energy \,$\hat E_{{p}i,{v}j}^{SP}$\, in
(\ref{W_ij^pv}) depends in general on the atomic configuration near
the \,$ij$\, bond. We describe this dependence by the pairwise
interaction model \cite{SF-07,VZh-12} and write this energy as
follows:
\begin{equation}
\hat E_{{p}i,{v}j}^{SP}=E_{h}^{p}+\sum_{\lambda l}
\Delta_{p,ij}^{\lambda l}n_l^{\lambda},\quad \Delta_{p,ij}^{\lambda
l}\,= \,(\varepsilon^{\lambda
l}_{p,ij}-\varepsilon^{hl}_{p,ij}).\label{E^SP_p}
\end{equation}
Here \,$E_{h}^{p}$\, is the saddle point energy  for a $p$-species
atom in the pure host metal, the parameter $\Delta^{\lambda
l}_{p,ij}$ (to be called the ``saddle-point interaction'') describes
changes in this energy due to a possible substitution of a host atom
in site \,$l$ by a $\lambda$-species solute atom, \, while
\,$\varepsilon^{\lambda l}_{p,ij}$ and $\varepsilon^{hl}_{p,ij}$ are
microscopic parameters which can be calculated using either $ab \ \
initio$ \cite{SF-07} or model \cite{VZh-12} calculations.

The most general expression for the probability \,$P$\, in
(\ref{dP/dt}) can be written as \cite{BV-98,Nastar-00}
\begin{eqnarray}
&P\{n_i^{\rho}\}=&\exp
\Big[\beta\Big(\Omega+\sum_{{\rho}i}\lambda_i^{\rho}n_i^{\rho}
-H_{int}-\hat{h}_{int}\Big)\Big],\label{P}\\
&\hat{h}_{int}=&\frac{1}{2}\sum_{\rho\sigma,ij}
h_{ij}^{\rho\sigma}n_i^{\rho}n_j^{\sigma}\nonumber\\
&&+\frac{1}{6}\sum_{\rho\sigma\tau,ijk}
h_{ijk}^{\rho\sigma\tau}n_i^{\rho}n_j^{\sigma}n_k^{\tau}+\ldots
\label{h_rho-sigma}
\end{eqnarray}
Here parameters \,$\lambda_i^{\rho}$\,  (which are both time- and
space-dependent, in general) can be called ``site chemical
potentials'' for an \,$\alpha$-species atom or a vacancy with
respect to a host atom. These parameters are related to the local
chemical potentials $\mu_i^{\rho}$ and $\mu_i^h$ as \cite{VZhKh-10}:
\begin{equation}
\lambda_i^{\rho}= (\mu_i^{\rho}- \mu_i^{h}).\label{lambda_i-def}
\end{equation}
Quantities $h_{ij...}^{\rho\sigma...}$ in (\ref{h_rho-sigma}) (to be
called ``effective interactions''
\hbox{\cite{Nastar-00,Nastar-05,Barbe-06})} describe
renormalizations of configurational interactions  (\ref{v_ij-def})
in the course of kinetic processes, and they can depend on both time
and space, too. Constant $\Omega$ is determined by normalization.

Multiplying Eq. (\ref{dP/dt}) by operators $n_i^{\rho}$ and summing
over  all configurations $\{n_j^{\lambda}\}$, we obtain equations
for the mean occupations of site (``local site fractions'')
\,$c_i^{\rho}=\langle n_i^{\rho}\rangle$:
\begin{equation}
dc_i^{\rho}/dt= \langle n_i^{\rho}\hat S\rangle \label{c_rho-dot}
\end{equation}
where\,$\langle(...)\rangle$  means averaging over distribution
(\ref{P}), e. g.:
\begin{equation}
c_i^{\rho}=\langle n_i^{\rho}\rangle =
\sum_{\{n_j^{\sigma}\}}n_i^{\rho}P\{n_j^{\sigma}\}. \label{c_rho}
\end{equation}

For simplicity, in this work we consider the case of presence in
(\ref{h_rho-sigma}) of only pairwise effective interactions
$h_{ij}^{\rho\sigma}$ which is sufficient for dilute alloys while.
non-pairwise effective interactions will be discussed elsewhere.
Then after some manipulations described in detail in \cite{VZh-12},
Eqs. (\ref{c_rho-dot}) can be written similarly  to Eqs.
\hbox{(I-28)-(I-34):}
\begin{eqnarray}
&dc_i^{\alpha}/dt=&\sum_{j(i)} \Big\langle \gamma_{\alpha v}
\hat{b}^{\alpha}_{ij}\Big\{\exp\,\Big[\beta \Big(\lambda_j^{\alpha}+
\lambda_i^v-h_{ji}^{\alpha v}\nonumber\\
&&-\sum_{\lambda
l}(h_{jl}^{\alpha\lambda}+h_{il}^{v\lambda})n_l^{\lambda} \Big)\Big]
-\{i\to j\}\Big\}\Big\rangle\nonumber\\
&dc_i^h/dt=&\sum_{j(i)} \Big\langle \gamma_{h v}
\hat{b}^h_{ij}\Big\{\exp\,\Big[\beta \Big(\lambda_i^v-\sum_{\lambda
l}h_{il}^{v\lambda}n_l^{\lambda} \Big)\Big]\nonumber\\
&&-\{i\to j\}\Big\}\Big\rangle.\label{dc^alpha,h-dt}
\end{eqnarray}
where we also correct some misprints made in \cite{VZh-12} and use
the identity $(c^v_i+ \sum_{\alpha}c^{\alpha}_i)=(1-c^h_i)$. In Eqs.
(\ref{dc^alpha,h-dt}), symbol \,$j(i)$\, means summation over sites
\,$j$\, being nearest neighbors of site \,$i$,\, and the factor
$\gamma_{pv}$ can be called ``the activation frequency'' for a $p\to
v$ exchange  in a pure host metal which can be written similarly to
(\ref{W_ij^pv}):
\begin{equation}
\gamma_{pv}=\omega_{pv}^{eff} \exp\,(-\beta E_{ac}^{pv}).
\label{gamma^pv}
\end{equation}
Here \,$\omega_{pv}^{eff}$\,  is the same as  in (\ref{omega_pv}),
while \,$E_{ac}^{pv}$\, is the effective activation energy which is
linearly expressed via the saddle point energy \,$E_{h}^{p}$\, in
(\ref{E^SP_p}) and couplings \,$V_{ij}^{pp'}$ \cite{VZh-12}. The
operator $\hat{b}^p_{ij}$ in (\ref{dc^alpha,h-dt}) is given by Eq.
(I-33):
\begin{equation}
\hat{b}^p_{ij}=n_i^hn_j^h\exp\Big[\sum_{\alpha
l}\beta(u_{il}^{\alpha}+u_{jl}^{\alpha})n_l^{\alpha}- \sum_{\alpha
l}\beta\Delta_{p,ij}^{\alpha l}n_l^{\alpha}\Big]\label{b_ij^p}
\end{equation}
where \,$\Delta_{p,ij}^{\alpha l}$\, is the same as in Eq.
(\ref{E^SP_p}), while parameters $u_{il}^{\alpha}$ (to be called
``kinetic interactions'' \cite{KSSV-11}) are expressed via
$V_{ij}^{pq}$ in (\ref{H^t}) as follows:\,
\begin{equation}
u^{\alpha}_{il}= (V_{il}^{h\alpha}-V_{il}^{hh}).\label{u^alpha_il}
\end{equation}
Eqs. (\ref{dc^alpha,h-dt})-(\ref{b_ij^p}) show that the operator
$\hat{b}^p_{ij}$ describes influence of neighboring solute atoms on
the probability of a p$i\rightleftharpoons {v}j$\, jump. Note that
the kinetic interaction $u_{il}^{\alpha}$ in (\ref{b_ij^p}) and
(\ref{u^alpha_il}) does not depend on the kind $p$ of a jumping
atom, unlike the saddle-point interaction $\Delta_{p,ij}^{\alpha l}$
in (\ref{E^SP_p}).

Using the operator identities
\begin{eqnarray}
\hskip-15mm&&n_l^{\alpha}n_l^{\beta}=n_l^{\alpha}\delta_{\alpha\beta},\quad
\exp(xn_l^{\alpha})=1+n_l^{\alpha}f(x)\label{e^(xn)=1+nf}
\end{eqnarray}
where $f(x)$ is $(e^x-1)$ and $\delta_{\alpha\beta}$ is the Kroneker
symbol, we can explicitly write the operator $\hat{b}^p_{ij}$
(\ref{b_ij^p}) as follows :
\begin{eqnarray}
&&\hat{b}^p_{ij}=n_i^hn_j^h\prod_l(1+f_{p\Delta,ij}^{\alpha
l}n_l^{\alpha}),\label{b_ij^p-product}\\
&&f_{p\Delta,ij}^{\alpha l}=[\exp(\beta\Delta_{p,ij}^{\alpha l}-
\beta u_{il}^{\alpha}-\beta u_{jl}^{\alpha})-1].\label{f_p-delta}
\end{eqnarray}

Finally, we make remarks on the difference between our formulation
of the master equation approach and that used by Nastar
\cite{Nastar-05}. First, our formulation is based on the explicit
expression (\ref{W_ij^pv}) for the inter-site atomic exchange
probability $W_{ij}^{pv}$. On the contrary, Nastar  treats this
$W_{ij}^{pv}$ as some unknown operator and estimates averages with
this operator for dilute alloys using some indirect considerations
rather than the direct calculations. For concentrated alloys, she
mentions ``difficulties'' to construct $W_{ij}^{pv}$ ``satisfying
the detailed balance principle'', while this principle is
identically obeyed for our expression (\ref{W_ij^pv}). As the
result, Ref. \cite{Nastar-05} does not contain  explicit equations
(\ref{dc^alpha,h-dt}) for time derivatives $dc^{\rho}_i/dt$ which
include, in particular, the vacancy activity factor
$\exp\,(\beta\lambda^v_i)$ discussed below. Second, the
above-mentioned elimination by Nastar et al.
\cite{Nastar-00,Nastar-05,Barbe-06} of vacancy occupation operators
$n_i^v$ [rather than operators $n_i^h$ in Eq. (\ref{n^h_i})] can
lead to difficulties in practical calculations, for example, in the
$ab \ initio$ calculations of interactions (\ref{v_ij-def}) by
standard methods \cite{SF-07}. Such difficulties are absent in our
formulation based on Eqs. (\ref{n^h_i})-(\ref{b_ij^p-product}).

\section{GENERAL EQUATIONS FOR ONSAGER COEFFICIENTS\label{Gen-eq-for-Onsager}}

\subsection{Method of calculations of Onsager coefficients
in the master equation
approach\label{Calc-Onsag-Master-eq-approach}}

The steady-state diffusion is commonly described in terms of Onsager
coefficients $L_{pq}$ which relate the atomic flux density ${\bf
J}_p$ to the chemical potential gradients $\nabla\mu_q$ supposed to
be small and constant \cite{Allnatt-93}. These chemical potentials
can be counted off the vacancy chemical potential $\mu_v$, and in
the cubic metals where diffusion is isotropic, Onsager relations can
be written as:
\begin{equation}
{\bf J}_p=\sum_qL_{pq}\nabla\mu_{qv} \label{J-nabla-mu}
\end{equation}
where $\mu_{qv}$ is $(\mu_{q}-\mu_{v})$. In a nonuniform alloy,
local values $\mu^{qv}_i$ are related to $\lambda_i^{\rho}$ defined
by Eqs. (\ref{lambda_i-def}) as follows:
\begin{equation}
\mu_i^{\alpha v}=(\lambda_i^{\alpha}-\lambda_i^v)\,,\qquad \mu_i^{h
v}=-\lambda_i^v\,. \label{mu^pv-def}
\end{equation}

Below we use the methods of calculations of Onsager coefficients
developed by Nastar et al. \cite{Nastar-00,Nastar-05,Barbe-06}. The
steady-state diffusion corresponds to a weakly non-uniform alloy for
which the local chemical potential difference
$\delta\lambda^{\rho}_{ji}= (\lambda_j^{\rho}-\lambda_i^{\rho})$ in
Eqs. (\ref{dc^alpha,h-dt}) is small, while effective interactions
$h_{ij}^{\alpha\rho}$  (for brevity, to be called also ``fields'')
are proportional to these differences. Linearizing  Eqs.
(\ref{dc^alpha,h-dt}) in $\delta\lambda^{\rho}_{ji}$ and
$h_{ij}^{\alpha\rho}$ and expressing $\delta\lambda^{\rho}_{ji}$ via
$\delta\mu^{pv}_{ji}=(\mu^{pv}_{j}-\mu^{pv}_{i})$ according to
(\ref{mu^pv-def}), we obtain:
\begin{eqnarray}
&dc_i^p/dt=&\beta\sum_{j(i)}\Big\langle \gamma_{pv}
\exp\,(\beta\lambda_{\alpha}+\beta\lambda_v)\hat{b}^p_{ij}\nonumber\\
&&\times\Big[\delta\mu^{pv}_{ji}+(h_{ij}^{pv}-h_{ji}^{pv})-
\sum_{\lambda l}(h_{il}^{v\lambda}
-h_{jl}^{v\lambda})n_l^{\lambda}\nonumber\\
&&+\sum_{\lambda l}
(h_{il}^{p\lambda}-h_{jl}^{p\lambda})n_l^{\lambda}
\Big]\Big\rangle.\label{dc^alpha,h/dt-lin}
\end{eqnarray}
Here and below, $\lambda_{\alpha}$ or $\lambda_v$ without a site
index $i$ or $j$ means  the equilibrium value of this chemical
potential, while averaging is made over the equilibrium distribution
$P$ described by Eq. (\ref{P}) with $\lambda^{\rho}_i
=\lambda_{\rho}$ and $\hat{h}_{int}=0$. In accordance with the
definition (\ref{h_rho-sigma}), fields $h^{p\lambda}_{ij}$ are
nonzero only when index $p$ corresponds to a solute atom $\alpha$,
while $h^{h\lambda}_{ij}=0$ [which is also illustrated by Eqs.
(\ref{dc^alpha,h-dt})]. For the given $j$, each term in the
right-hand side of (\ref{dc^alpha,h/dt-lin}) has evidently the
meaning of an atomic flux $J^p_{j\to i}$ through bond $ij$. It
enables us to write the linear relation between these fluxes and
quantities $\delta\mu^{pv}_{ji}$ and $h_{ij}^{\alpha\rho}$ in
(\ref{dc^alpha,h/dt-lin}). It was also noted in
\cite{Nastar-00,Nastar-05} that for the steady-state diffusion when
the left-hand side of Eqs. (\ref{dc^alpha,h/dt-lin}) vanishes,
effective interactions $h_{ij}^{\alpha\rho}$ are antisymmetric in
indices $i$ and $j$:
\begin{equation}
h_{ji}^{\alpha v}=-h_{ij}^{\alpha v},\qquad
h_{ji}^{\alpha\beta}=-h_{ij}^{\alpha\beta}.\label{h_ij-minus_h_ji}
\end{equation}
Denoting also site $i$ by index ``0'' and site $j$ by index ``1'',
we can write the above-mentioned fluxes $J^p_{0\to 1}$ as follows:
\begin{eqnarray}
& J^{p}_{0\to 1}=&-\beta\Big[\overline{w}_{p}
(\delta\mu_{\alpha v} +2h_1^{\alpha v})\nonumber\\
&&-\sum_{\lambda l}l_{p}^{\lambda l}
(h_{0l}^{v\lambda}-h_{1l}^{v\lambda}- h_{0l}^{p\lambda}+
h_{1l}^{p\lambda})\Big]\label{J^alpha,h}
\end{eqnarray}
where $\delta\mu_{pv}$ is $(\mu^{pv}_1-\mu^{pv}_0)$,\, $h_1^{\alpha
v}$ is the nearest-neighbor effective interaction, and
$\overline{w}_p$ and $l_p^{\lambda l}$ are statistical averages:
\begin{equation}
\overline{w}_p=\langle\hat{w}_{01}^p\rangle,\qquad l_p^{\lambda
l}=\langle\hat{w}_{01}^pn^{\lambda}_l\rangle. \label{overline-w^p}
\end{equation}
Here the operator $\hat{w}_{01}^p$ is the product of the operator
$\hat{b}_{01}^{p}$ given by Eq. (\ref{b_ij^p}) or
(\ref{b_ij^p-product}) and the constant factor $\Gamma_p$ which
enters into Eqs. (\ref{dc^alpha,h/dt-lin}):
\begin{eqnarray}
\hskip-10mm&&\hat{w}_{01}^p=\Gamma_p\,\hat{b}_{01}^{p},\label{hat-w}\\
\hskip-10mm&&\Gamma_{\alpha}=\gamma_{\alpha
v}\exp\,(\beta\lambda_{\alpha}+\beta\lambda_v),\quad\Gamma_h=\gamma_{hv}\exp\,(\beta\lambda_v).
\label{Gamma_p}
\end{eqnarray}
Fields $h_{ij}^{\alpha\rho}$ in Eqs. (\ref{J^alpha,h}) can be found
from the stationarity condition for two-site averages
\cite{Nastar-00,Nastar-05}:
\begin{equation}
{d\over dt}\langle n^{\alpha}_0n^p_j\rangle=0 \label{d/dt(nn)=0}
\end{equation}
which yields the following equations for $h_{ij}^{\alpha\rho}$
\cite{Nastar-00,Nastar-05,BV-98}:
\begin{eqnarray}
\hskip-5mm&&\sum_{k\neq 0\neq j}\Big[m_{\alpha,0k}^{pj}\Big(
\delta\mu_{k0}^{\alpha v} +2h_{0k}^{\alpha v}\Big)-\sum_{\lambda
l}t_{\alpha,0k}^{pj,\lambda l}\Big
(h_{0l}^{v\lambda}- h_{kl}^{v\lambda}\nonumber\\
&&\hskip5mm- h_{0l}^{\alpha\lambda}+ h_{kl}^{\alpha\lambda}\Big)
+m_{h,jk}^{\alpha 0}(\delta\mu_{kj}^{pv}+2h^{pv}_{jk})\nonumber\\
&&\hskip5mm-\sum_{\lambda l}t_{h,jk}^{\alpha 0,\lambda l}\Big
(h_{jl}^{v\lambda}- h_{kl}^{v\lambda}- h_{jl}^{p\lambda}+
h_{kl}^{p\lambda}\Big)\Big]=0\, \label{m-t-gen}
\end{eqnarray}
where
\begin{equation}
m_{p,ik}^{qj}=\langle\hat{w}_{ik}^pn^q_j\rangle,\qquad
t_{p,ik}^{qj,\lambda l}=
\langle\hat{w}_{ik}^pn^q_jn^{\lambda}_l\rangle.\label{m-t-gen-3}
\end{equation}

Following Nastar \cite{Nastar-05}, we consider diffusion along
$z$-axis of an FCC alloy when chemical potentials
$\mu^p_i=\mu^p({\bf R}_i)$ depend only on $z_i$. Let us denote
positions of sites $0$ and $1$ in Eqs. (\ref{J^alpha,h}) as ${\bf
R}_0=(0,0,0)$ and ${\bf R}_1=(0,a_0/2,a_0/2)$ where $a_0$ is the FCC
lattice constant, while different sites near the bond $(0,1)$ are
numbered as shown in Fig. 1. Quantity $\delta\mu_{pv}$ in Eqs.
(\ref{J^alpha,h}) is the difference of chemical potentials between
neighboring atomic planes along $z$ axis:
$\delta\mu_{pv}=\mu_{pv}(0,0,a_0/2)-\mu_{pv}(0,0,0)$. The field
$h_{0l}^{\rho\lambda}= h^{\rho\lambda}({\bf R}_{0l})$ does not
change under rotations of vector ${\bf
R}_{0l}=(x_{0l},y_{0l},z_{0l})$ around $z$-axis,  and this field
changes  its sign under reflection with respect to $(x,y)$-plane:
$h^{\rho\lambda}(x_{0l},y_{0l},-z_{0l})=
-h^{\rho\lambda}(x_{0l},y_{0l},z_{0l})$. For brevity, we denote the
set of crystallographically equivalent sites with the same positive
value $z_{0l_n}>0$ as $l_n^+$, the similar set with the negative
value $z_{0l_n}=-z_{l_n}$, as $l_n^-$, and the fields
$h^{\rho\lambda}({\bf R}_{l_n^+})$ or $h^{\rho\lambda}({\bf
R}_{l_n^-})$ which correspond to the set of sites $l_n^+$ or
$l_n^-$, as $h_n^{\rho\lambda}$ or $(-h_n^{\rho\lambda})$. Index $n$
which numbers different sets of equivalent sites, $l_n^+$ and
$l_n^-$, is supposed to increase with the distance $|{\bf R}_{0l}|$,
and for a given  $|{\bf R}_{0l}|$, it increases with the $z_{0l}$
value. Thus $n=1$ corresponds to the nearest-neighbor field
$h_1=h({\bf R}_{01})$, and Eqs. (\ref{J^alpha,h}) can be concisely
written as:
\begin{eqnarray}
& J^p_{0\to 1}=&-\beta\Big[\overline{w}_p(\delta\mu_{\alpha v}
+2h_1^{pv})\nonumber\\
&&+\sum_{\lambda}\sum_{n=1}^{n_{max}}l_{p,n}^{\lambda}
(h_{n}^{\lambda v}-h_{n}^{\lambda p})\Big]. \label{J^alpha,h-l_n}
\end{eqnarray}
Here  $n_{max}$ is the maximum number of fields $h_n^{\rho\lambda}$
taken into account, and increase of $n_{max}$ corresponds to a more
accurate description of vacancy correlation effects
\cite{Nastar-00}. Coefficients $l_{p,n}^{\lambda}$ in
(\ref{J^alpha,h-l_n}) are defined as follows:
\begin{equation}
l_{p,n}^{\lambda}=\sum_{l_n^+,\,l_n^-}\langle\hat{w}_{01}^p
(n_{l_n^+}-n_{l_n^-}-n_{1,l_n^+}+n_{1,l_n^-})^{\lambda}\rangle.\label{l_p,n}
\end{equation}
Here  index $\lambda$ at brackets means that it should be put at
each term within brackets, e. g.
$(n_{l_n^+}+\ldots)^{\lambda}=(n_{l_n^+}^{\lambda}+\ldots)$, and the
following notation is used:
\begin{eqnarray}
&& n_{l_n^{\pm}}=n({\bf R}_{l^{\pm}_n}),\quad n_{1,l_n^{\pm}}=n({\bf
R}_{l^{\pm}_n}+{\bf R}_{1}).\label{n_l^pm}
\end{eqnarray}

Employing the same notation as in
(\ref{J^alpha,h-l_n})-(\ref{n_l^pm}), we  can concisely write Eqs.
(\ref{m-t-gen}) and (\ref{m-t-gen-3}) similarly to
(\ref{J^alpha,h-l_n}):
\begin{eqnarray}
\hskip-10mm&&m_{\alpha,n}^{p}(\delta\mu_{\alpha v}+2h_1^{\alpha
v})-m_{p,n}^{\alpha}(\delta\mu_{pv}
2h_1^{pv})+\nonumber\\
\hskip-10mm&&\hskip10mm+\sum_{\lambda}\sum_{m=1}^{n_{max}}
\Big[(t_{\alpha,nm}^{p\lambda}
-{t}_{p,nm}^{\alpha\lambda})h_{m}^{\lambda v}\nonumber\\
\hskip-10mm&&\hskip10mm- t_{\alpha,nm}^{p\lambda}
h_{m}^{\lambda\alpha}+ t_{p,nm}^{\alpha\lambda} h_{m}^{\lambda
p}\Big]=0\, \label{m-t-expl}
\end{eqnarray}
where coefficients  $t_{p,nm}^{q\lambda}$ and $m_{p,n}^q$ are
defined as follows:
\begin{eqnarray}
\hskip-2mm&t_{p,nm}^{q\lambda}=&\sum_{k=1}^{12}\sum_{l_m^+,\,l_m^-}
\langle\hat{w}_{0k}^pn_{n,1}^q
(n_{l_m^+}-n_{l_m^-}-n_{k,l_m^+}+n_{k,l_m^-})^{\lambda}\nonumber\\
\hskip-2mm&m_{p,n}^q=&\sum_{k=1}^4\langle(\hat{w}_{0k}^p-
\hat{w}^p_{0,k+4})n_{n,1}^q\rangle\, .\label{m,t_p^q,n}
\end{eqnarray}
Here $n_{l_m^{+}}$ and  $n_{l_m^{-}}$ are the same as in
(\ref{l_p,n}); the operator $n_{n,1}^q=n^q({\bf R}_{n,1})$
corresponds to the vector ${\bf R}_{n,1}$  chosen as ``the first
one'' in the set of vectors  ${\bf R}_{l_n^+}$; $n_{k,l_n^{\pm}}$
defined similarly to $n_{1,l_n^{\pm}}$ in Eq. (\ref{n_l^pm}) is
$n({\bf R}_{l^{\pm}_n}+{\bf R}_{k})$;  and we took unto account the
symmetry or antisymmetry of each average  in (\ref{m-t-expl}) with
respect to reflections ${\bf R}_{n,1}\to (-{\bf R}_{n,1})$.

Eqs.  (\ref{m-t-expl})  enable us to express all fields
$h_n^{\alpha\rho}$ as linear combinations of $\delta\mu_{qv}$. Then
substitution of these expressions into Eqs. (\ref{J^alpha,h-l_n})
yields the linear relation between the flux $J^{p}_{0\to 1}$ and
differences $\delta\mu_{pv}$:
\begin{equation}
J^{p}_{0\to 1}=\sum_qA_{pq}\delta\mu_{qv} \label{A_pq}
\end{equation}
where parameters $A_{pq}$ are some functions of coefficients
$l_{p,n}^{\lambda}$, ${m}_{p,n}^q$ and $t_{p,n}^{q\lambda}$ in Eqs.
(\ref{m-t-expl}). To relate  parameters $A_{pq}$ in (\ref{A_pq}) to
Onsager coefficients $L_{pq}$ in (\ref{J-nabla-mu}) we note that the
flux density $J_p$ along $z$ axis can be found as the ratio of the
total flux through one site lying in the plane (0,0,0) to the area
$S=a_0^2/2$ corresponding to each site in that plane, while the
difference $\delta\mu_{pv}$ in Eq. (\ref{A_pq}) is simply expressed
via $\nabla\mu_{qv}= (0,0,d\mu_{qv}/dz)$:
\begin{eqnarray}
&&J_p=4 J^{p}_{0\to 1}/S=8J^{p}_{i\to
j}/a_0^2\,,\label{J_p-J^p_01}\\
&&\delta\mu_{pv}=(d\mu_{pv}/dz)a_0/2\label{delta_mu-nabla_mu}
\end{eqnarray}
Substituting these relations into (\ref{A_pq}) and comparing the
result with a $z$-component of Eq.  (\ref{J-nabla-mu}), we find:
\begin{equation}
L_{pq}=-4A_{pq}/a_0=-na_0^2A_{pq}\label{L-A_pq}
\end{equation}
where $n=4/a_0^3$ is the atomic density in the FCC lattice.

\subsection{Model of the nearest-neighbor kinetic and saddle-point
interactions}

Below we consider the model when both the saddle-point and  kinetic
interactions,  $\Delta_{p,ij}^{\lambda l}$ and $u_{il}^{\lambda}$ in
Eqs. (\ref{E^SP_p}), (\ref{b_ij^p}), (\ref{b_ij^p-product}) and
(\ref{f_p-delta}), are nonzero only for the nearest-neighbors. This
corresponds to the standard ``five-frequency model'' for FCC alloys
\cite{Lidiard-55,Lidiard-60,Manning-64,Howard-Man-67,Bocquet-74,
LeClaire-78,Ishioka-84,Allnatt-93, Bocquet-96}. For this model, the
operator $\hat{w}_{01}^{p}$ in Eqs. (\ref{b_ij^p-product}) and
(\ref{hat-w}) takes the form:
\begin{equation}
\hat{w}_{01}^{p}=\Gamma_p n_0^hn_1^h\prod_{l}
\Big(1+\sum_{\mu}n_l^{\mu}f_{p\Delta}^{\mu}\Big) \prod_{m}
\Big(1+\sum_{\nu}n_{m}^{\nu}f_{u}^{\nu}\Big) \label{w_01^p}
\end{equation}
where sites $l$ and $m$ are numbered as shown in Fig. 1. In this
figure, sites with positions  ${\bf R}_k$ for $k$ between 1 and 12
correspond to the nearest neighbors of site ``0'' positioned at
${\bf R}_0=0$, while sites positioned at ${\bf R}_{\bar{k}}\equiv
{\bf R}_{1,k}=({\bf R}_1+{\bf R}_k)$ correspond to the nearest
neighbors of site ``1'' with ${\bf R}_1=(0,a_0/2,a_0/2)$. The
relations
\begin{equation}
\bar{7}=0,\quad \bar{6}=9,\quad \bar{8}={12}, \quad
\overline{10}=2,\quad \overline{11}=4 \label{bar-k-k}
\end{equation}
are also taken into account.


%
\begin{figure}
\caption{Bond (0,1) in the FCC lattice and its nearest neighbors,
sites $k$ and $\bar{k}$ discussed in the text.
\label{sites-k,bar-k}}
\end{figure}

\begin{figure}
\caption{(color online) Schematic representation of bonds of the
type $(h,h)$, $(h,\Delta)$, $(\Delta,\Delta)$, $(\Delta, u)$ and
$(u,u)$ described in the text. Seven bonds $(0,k)$ and seven bonds
$(1,\bar{k})$ which belong to the type $(h,u)$ are not shown for
clarity of figure. \label{bonds}}
\end{figure}

In Eq. (\ref{w_01^p}), index $l$ in the first product takes four
values: 2, 4, 9 or 12 which correspond to the nearest neighbors of
bond (0,1), i. e.  of both site 0 and site 1. Index ${m}$ in the
second product corresponds to the nearest neighbors of only one of
these sites, site 0 or site 1, and takes fourteen values: 3, 5, 6,
7, 8, 10, 11, $\bar 1, \bar 2, \bar 3, \bar 4,\bar 5, \bar 9$ or
$\overline{12}$. Quantity $f_{p\Delta}^{\mu}$ or $f_{u}^{\nu}$ in
Eq. (\ref{w_01^p})  is the Mayer function which, according to Eq.
(\ref{f_p-delta}), corresponds to the sum of non-zero contributions
of the saddle-point interaction (\ref{E^SP_p}) and the kinetic
interaction  (\ref{u^alpha_il}):
\begin{equation}
f_{p\Delta}^{\mu}=\exp\,[\beta(2u_1^{\mu}-\Delta_{p}^{\mu})]-1\,,
\quad f_{u}^{\nu}=\exp\,(\beta u_1^{\nu})-1 \label{f_Delta,u}
\end{equation}
where $u_1^{\nu}$ is the nearest-neighbor kinetic interaction.

The vacancy correlation effects in concentrated alloys will be
described using two different approximations:

(i) The simplest ``Lidiard-Le Claire'' approximation which supposes
that a vacancy that leaves the first neighbor shell of a solute atom
does not return \cite{Allnatt-93}. It corresponds to the
nearest-neighbor effective interaction: $h_n=\delta_{n,1}h_1$
\cite{Nastar-05} and will be called the ``nearest-neighbor-jump
approximation'' (NNJA). For the NNJA, Eqs.
(\ref{J^alpha,h-l_n})-(\ref{m-t-expl}) include only terms with $n=1$
and $m=1$, and  Eqs. (\ref{l_p,n}) and (\ref{m,t_p^q,n}) take the
form:
\begin{eqnarray}
\hskip-10mm&l_{p,1}^{\lambda}=&\sum_{k=1}^4\langle\hat{w}_{01}^p(n_k-n_{k+4}-
n_{1,k}+n_{1,k+4})\rangle\,,\label{l_p^lambda,nn-jump}\\
\hskip-10mm&m_{p,1}^q=&\sum_{k=1}^4\langle(\hat{w}_{0k}^p-
\hat{w}^p_{0,k+4})n_{1}^q\rangle\,,\label{m_p^q,nn-jump}\\
\hskip-10mm&t_{p,11}^{q\lambda}=&\sum_{k=1}^{12}\langle(\hat{w}_{0k}^pn_{1}^q\sum_{l=1}^4
(n_{l}-n_{l+4}-n_{k,l}+n_{k,l+4})^{\lambda}\rangle\,.\label{t_p^q,nn-jump}
\end{eqnarray}

(ii) The more refined approximation suggested by Bocquet
\cite{Bocquet-74} which neglects the probability of return of a
vacancy which leaves the second shell of neighbors, to be called
``the second-shell-jump'' approximation (SSJA). It describes the
vacancy correlation effects with the accuracy of the order of
percents \cite{Bocquet-74} sufficient for the most of applications.
In Eqs. (\ref{l_p,n})-(\ref{m-t-expl}), SSJA corresponds to
$n_{max}=5$, that is, to the presence of five fields $h_n$ with the
following vectors ${\bf R}_{n,1}$ in Eq. (\ref{m-t-expl}) (in
$a_0/2$ units):
\begin{eqnarray}
&&{\bf R}_{1,1}=(0,1,1),\quad {\bf R}_{2,1}=(0,0,2),\quad {\bf
R}_{3,1}=(1,2,1),\nonumber\\
&& {\bf R}_{4,1}=(1,1,2),\quad {\bf R}_{5,1}=(0,0,2) \label{R_n,1}
\end{eqnarray}
while the set $l_n^+$ of vectors ${\bf R}_{l_n^+}$ in Eqs.
(\ref{l_p,n}) and (\ref{m,t_p^q,n})  for $n$ equal to 1, 2, 3, 4 and
5 includes 6, 1, 8, 4 and 4 vectors ${\bf R}_{l_n^+}$, respectively.

Therefore, to find atomic fluxes $J^{p}_{0\to 1}$ in Eqs.
(\ref{J^alpha,h-l_n}) we should  calculate statistical averages of
three different types: quantities $\overline{w}_p=
\langle\hat{w}_{01}^p\rangle$ in Eq. (\ref{w_01^p}); quantities
$l_{p,n}^{\lambda}$ and ${m}_{p,n}^q$ in Eqs. (\ref{l_p,n}) and
(\ref{m,t_p^q,n})  which include ``one-site'' averages
$\langle\hat{w}_{01}^pn^{\lambda}_i\rangle$; and quantities
$t_{p,nm}^{q\lambda}$ in (\ref{m,t_p^q,n}) which include
``two-site'' averages
$\langle\hat{w}_{01}^pn^q_in^{\lambda}_j\rangle$.

\section{CALCULATIONS OF STATISTICAL AVERAGES\label{Calc-averages-gen}}.

\subsection{Exact relations\label{Exact-relatons}}

Before to discuss methods of calculations of averages
$\overline{w}_p$, $l_{p,n}^{\lambda}$, ${m}_{p,n}^q$ and
$t_{p,nm}^{q\lambda}$ in Eqs.
(\ref{J^alpha,h-l_n})-(\ref{m,t_p^q,n}) we consider some exact
relations which follow either from definitions of these averages or
from the crystal symmetry.

First, we note that according to definitions (\ref{overline-w^p}),
(\ref{l_p,n}),  (\ref{m,t_p^q,n}), each such average is proportional
to the factor $\exp\,(\beta\lambda_v)$, that is, to the reduced
thermodynamic activity coefficient $a_v$ for vacancies defined by
Eqs. (\ref{lambda_v,alpha})-(\ref{a_v,a_alpha}) below. This factor
enters into the coefficient $\Gamma_p$ in Eqs. (\ref{hat-w}) and
(\ref{Gamma_p}) and is determined by the vacancy-solute interactions
$v_{v\alpha}$. Therefore, at nonzero solute concentrations
$c_{\alpha}$ these $v_{v\alpha}$ affect all diffusion coefficients,
contrary to the commonly accepted ideas
\cite{Lidiard-60,Manning-64,Howard-Man-67,Bocquet-74,LeClaire-78,Ishioka-84,Allnatt-93,
Bocquet-96,Nastar-05}.

Second, we note two operator identities which are useful for
calculations of one-site or two-site averages, i. e. quantities
$l_{p,n}^{\lambda}$, ${m}_{p,n}^q$ or $t_{p,nm}^{q\lambda}$ in Eqs.
(\ref{J^alpha,h-l_n})-(\ref{m-t-expl}). These identities include the
product of the operator $n_i^q$ and one of factors in two last
products in Eq. (\ref{w_01^p}):
\begin{eqnarray}
&&n_i^q(1+\sum_{\lambda}n_i^{\lambda}f_{p\Delta}^{\lambda})=
n_i^qe_{p\Delta}^q,\nonumber\\
&&n_i^q(1+\sum_{\lambda}n_i^{\lambda}f_{u}^{\lambda})= n_i^qe_{u}^q
\label{identities-for-nu_i,ij}
\end{eqnarray}
where we denote for brevity:
\begin{equation}
e_{p\Delta}^q=\exp\,[\beta(2u_1^{q}-\Delta_{p}^{q})]\,, \qquad
e_{u}^q=\exp\,(\beta u_1^q)\,. \label{e-exp}
\end{equation}
Note that when index $q$ in Eqs. (\ref{identities-for-nu_i,ij})
corresponds to a host atom: $q=h$, the factor $e_{p\Delta}^h$ or
$e_{u}^h$ in (\ref{identities-for-nu_i,ij}) is unity:
\begin{equation}
e_{p\Delta}^h=e_{u}^h=1 \label{e-h-1}
\end{equation}
as the product $n_i^hn_i^{\lambda}$ in
(\ref{identities-for-nu_i,ij}) is zero. Eqs.
(\ref{identities-for-nu_i,ij}) imply, for example, that  in Eq.
(\ref{l_p^lambda,nn-jump}) for $l_{p}^{\lambda}$, the product
$(1+\sum_{\mu}n_2^{\mu}f_{p\Delta}^{\mu})n_2^{\lambda}$ in the
operator $\hat{w}_{01}^pn_2^{\lambda}$ is reduced to
$e_{p\Delta}^{\lambda}n_2^{\lambda}$, while the product
$(1+\sum_{\mu}n_6^{\mu}f_u^{\mu})n_6^{\lambda}$ is reduced to
$e_{u}^{\lambda}n_6^{\lambda}$. It simplifies calculations of
averages.

Third, we consider the crystal symmetry relations for one-site and
two-site averages, denoted as $\nu_{pi}^q$ and
$\nu_{p,ij}^{q\lambda}$:
\begin{equation}
\nu_{pi}^q=\langle\hat{w}_{01}^{p}n_i^q\rangle,\quad
\nu_{p,ij}^{q\lambda}=
\langle\hat{w}_{01}^{p}n_i^qn_j^{\lambda}\rangle\,.
\label{nu_i,ij-def}
\end{equation}
These relations can be conveniently discussed using Figs. 1 and 2
which illustrate the crystal symmetry of different sites near the
(0,1) bond for an inter-site jump $p\leftrightharpoons v$. These
sites can be divided into three groups: (i) sites 0 and $1\equiv\bar
0$, to be called ``sites $h$'' as occupation of these sites is
described in Eq. (\ref{w_01^p}) by the operators $n^h_0$ and
$n^h_1$; (ii) sites 2, 4, 9 and 12 being the nearest neighbors of
both sites 0 and site 1, to be called ``sites $\Delta$'' as the
occupation operator $n^{\lambda}_l$ for each of these sites enters
into Eq. (\ref{w_01^p}) with the factor $f_{p\Delta}^{\lambda}$;
(iii) the rest nearest neighbors of site 0 or site 1, that is, sites
3, 5, 6, 7, 8, 10, 11, and $\bar 1, \bar 2, \bar 3, \bar 4, \bar 5,
\bar 9$, $\overline{12}$, to be called ``sites $u$'' as the operator
$n^{\mu}_{m_1}$ or $n_{m_2}^{\nu}$ for these sites enters into Eq.
(\ref{w_01^p}) with the factor $f_{u}^{\mu}$ or $f_{u}^{\nu}$. The
sites $u$  can also be divided into three groups of the different
topology illustrated by Fig. 2: (i) the ``vertex'' sites 3, $\bar
3$, 5 and $\bar 5$, to be called ``sites $v$'', (ii) the ``side''
sites 6, 8, 10, 11, $\bar 2, \bar 4, \bar 9$ and $\overline{12}$, to
be called ``sites $s$'', and (iii) the ``central'' sites 7 and $\bar
1$, to be called ``sites $c$''. Different types of this site
symmetry  will be denoted by symbol $\xi$ which takes values
$\Delta$ and $u$ or, for a more detailed description, $\Delta$, $v$,
$s$ and $c$.

\vskip3mm

\vbox{\noindent TABLE I. Changes of positions of lattice sites under
rotation of the FCC lattice that transforms bond $(0,k)$ into bond
(0,1). \vskip3mm \hskip-4mm {\begin{tabular}{|c|c|cccccccccccc|}
\hline&Components &&&&&&&&&&&&\cr \ \ \ $k$ \ \ \ &of vector ${\bf
R}$&\multicolumn{11}{c}{\hskip7mm Position of sites}&\cr\hline
&&&&&&&&&&&&&\cr 1 &$(x,y,z)$&1&2&3&4&5&6&7&8&9&10&11&12\cr
&&&&&&&&&&&&&\cr 2 &$(-y,x,z)$ &
4&1&2&3&8&5&6&7&12&9&10&11\cr&&&&&&&&&&&&&\cr 3
&$(-x,-y,z)$&3&4&1&2&7&8&5&6&11&12&9&10\cr&&&&&&&&&&&&&\cr 4
&$(y,-x,z)$&2&3&4&1&6&7&8&5&10&11&12&9\cr&&&&&&&&&&&&&\cr 5
&$(x,-z,y)$&3&10&7&11&1&9&5&12&2&6&8&4\cr&&&&&&&&&&&&&\cr 6
&$(-y,-z,x)$&11&3&10&7&12&1&9&5&4&2&6&8\cr&&&&&&&&&&&&&\cr 7
&$(x,-y,-z)$&7&6&5&8&3&2&1&4&10&9&12&11\cr&&&&&&&&&&&&&\cr 8
&$(y,-z,-x)$&10&7&11&3&9&5&12&1&6&8&4&2\cr&&&&&&&&&&&&&\cr 9
&$(-z,y,x)$&12&4&11&8&9&2&10&6&1&3&7&5\cr&&&&&&&&&&&&&\cr 10
&$(-z,x,-y)$&8&12&4&11&6&9&2&10&5&1&3&7\cr&&&&&&&&&&&&&\cr 11
&$(z,-x,-y)$&6&10&2&9&8&11&4&12&7&3&1&5\cr&&&&&&&&&&&&&\cr
12 &$(z,y,-x)$&9&6&10&2&12&8&11&4&5&7&3&1\\
\hline
\end{tabular}}}

\vskip3mm

The above-discussed symmetry relations can be used to simplify Eq.
(\ref{l_p^lambda,nn-jump}) for $l_{p,1}^{\lambda}$ which is
originally written as
\begin{eqnarray}
\hskip-5mm&&l_{p,1}^{\lambda}=\langle\hat{w}_{01}^{p}[(n_1+n_2+n_3+n_4
-n_5-n_6-n_7-n_8)^{\lambda}\nonumber\\
\hskip-5mm&&-(n_{\bar 1}+n_{\bar 2}+n_{\bar 3}+n_{\bar 4}-n_{\bar
5}-n_{\bar 6}-n_{\bar 7}-n_{\bar 8})^{\lambda}]\rangle.
\label{l_p-expl-1}
\end{eqnarray}
First, three last terms in the second brackets can be rewritten
according to Eq. (\ref{bar-k-k}). Second, terms with $n^{\lambda}_0$
and $n_1^{\lambda}$ in (\ref{l_p-expl-1})  vanish as the operator
$\hat{w}_{01}^{p}$ (\ref{w_01^p}) includes factors $n_0^h$ and
$n_1^h$ while $n^h_in^{\lambda}_i=0$. Thus we obtain:
\begin{eqnarray}
\hskip-5mm&&l_{p,1}^{\lambda}=\langle\hat{w}_{01}^{p}[(n_2+n_4+n_9+n_{12})^{\lambda}
+(n_3-n_5\nonumber\\
\hskip-5mm&&-n_{\bar 3}+n_{\bar 5}-n_6-n_8-n_{\bar 2}-n_{\bar
4})-n_7-n_{\bar 1})^{\lambda}]\rangle. \label{l_p-expl-2}
\end{eqnarray}
Figs. 1 and 2 show that four $\Delta$-sites, 2, 4, 9, 12, as well as
four $v$-sites 3, 5, $\bar 3$, $\bar 5$, eight $s$-sites 6, 8, 10,
11, $\bar 2$,  $\bar 4$, $\bar 9$, $\overline{12}$, and two
$c$-sites, 7 and $\bar 1$, are equivalent to each other. Therefore,
Eq. (\ref{l_p-expl-2}) includes only three different terms:
\begin{equation}
l_{p,1}^{\lambda}=(4\nu_{p\Delta}^{\lambda}-4\nu_{ps}^{\lambda}-2\nu_{pc}^{\lambda})
\label{l_p-expl-3}
\end{equation}
where $\nu_{p\xi}^{\lambda}$ means the one-site average
$\nu_{pi}^{\lambda}$ (\ref{nu_i,ij-def}) for a site $i$ of the
symmetry $\xi$:
\begin{equation}
\nu_{p\Delta}^{\lambda}=\langle\hat{w}_{01}^{p}n_2^{\lambda}\rangle,\quad
\nu_{ps}^{\lambda}=\langle\hat{w}_{01}^{p}n_6^{\lambda}\rangle,\quad
\nu_{pc}^{\lambda}=\langle\hat{w}_{01}^{p}n_7^{\lambda}\rangle .
\label{l_p-expl-4}
\end{equation}

Expressions (\ref{m_p^q,nn-jump}) and (\ref{t_p^q,nn-jump}) for
$m_{p,1}^q$ and $t_{p,11}^{q\lambda}$ include operators
$\hat{w}_{0k}^{p}$ which describe atomic jumps along bonds $(0,k)$
rather than along the bond (0,1) considered above. To use the
above-discussed symmetry relations, we can employ the rotation of
the FCC lattice which transforms bond $(0,k)$ into the (0,1) one.
Table I shows changes of the positions of different sites under such
rotations.

Using Table I we can write $m_{p,1}^q$ in (\ref{m_p^q,nn-jump}) as
\begin{eqnarray}
&m_{p,1}^q=&\langle\hat{w}_{01}^{p}(n_4+n_3+n_2-n_3
-n_{11}-n_7-n_{10})^q\rangle\nonumber\\
&&= (2\nu_{p\Delta}^q-2\nu_{ps}^q-\nu_{pc}^q). \label{m_p^q-expl}
\end{eqnarray}
It implies:
\begin{equation}
l_{p,1}^q=2\,m_{p,1}^q \label{m_p^q-l_p^q}
\end{equation}
where we use the same considerations and notation as in Eqs.
(\ref{l_p-expl-1})-(\ref{l_p-expl-4}), while index $q$ corresponds
to either a solute atom $\lambda$ or a host atom $h$.

The similar methods can be used to explicitly write the average
$t_{p,11}^{q\lambda}$ in  (\ref{t_p^q,nn-jump}). It can be written
as the sum of two terms, the ``one-site'' and the ``two-site'' one:
\begin{equation}
t_{p,11}^{q\lambda}=t_{1p}^{q\lambda}+t_{2p}^{q\lambda}.
\label{t-t_1,2}
\end{equation}
The one-site term $t_{1p}^{q\lambda}$ has the form similar to
(\ref{l_p-expl-3}):
\begin{equation}
t_{1p}^{q\lambda}=\delta_{q\lambda}\Big(2\nu_{p\Delta}^{\lambda}+2\nu_{pv}^{\lambda}
+4\nu_{ps}^{\lambda}+\nu_{pc}^{\lambda}\Big) \label{t_1p}
\end{equation}
where $\nu_{pv}^{\lambda}$ is defined similarly to other
$\nu_{p\xi}^{\lambda}$ in (\ref{l_p-expl-4}):
\begin{equation}
\nu_{pv}^{\lambda}=\langle\hat{w}_{01}^{p}n_3^{\lambda}\rangle.
\label{l_pv^lambda}
\end{equation}
The two-site term $t_{2p}^{q\lambda}$ in (\ref{t-t_1,2}) includes 21
non-equivalent averages $\nu_{p,ij}^{q\lambda}$ which can be grouped
into terms $t_{p,\xi\xi'}^{q\lambda}$ corresponding to symmetries
$\xi$ and $\xi'$ of sites $i$ and $j$:
\begin{equation}
t_{2p}^{q\lambda}=\sum_{\xi,\xi'}t_{p,\xi\xi'}^{q\lambda}
\label{t_2p-sum-xi,xi'}
\end{equation}
where both $\xi$ and $\xi'$ takes the value $\Delta$, $v$, $s$ or
$c$. The non-zero terms $t_{p,\xi\xi'}^{q\lambda}$ in
(\ref{t_2p-sum-xi,xi'}) can be written as follows:
\begin{eqnarray}
\hskip-10mm&&t_{p,\Delta\Delta}^{q\lambda}=
(4\nu_{2,4}+2\nu_{2,9})_p^{q\lambda},\nonumber\\
\hskip-10mm&& t_{p,\Delta
s}^{q\lambda}=-2(\nu_{2,6}+\nu_{2,8}+\nu_{2,10}+\nu_{2,11})^{q\lambda}_p,\nonumber\\
\hskip-10mm&&t_{p,\Delta c}^{q\lambda}=-4(\nu_{2,7})_p^{q\lambda},
\quad t_{p,vv}^{q\lambda}=2(\nu_{3,{\bar 5}}-\nu_{3,5}-\nu_{3,{\bar 3}})^{q\lambda}_p,\nonumber\\
\hskip-10mm&&t_{p,vs}^{q\lambda}=
2(\nu_{3,10}-\nu_{3,6}-\nu_{3,{\bar 2}}+\nu_{3,{\bar 9}})^{q\lambda}_p,\nonumber\\
\hskip-10mm&&t_{p,ss}^{q\lambda}=
2(\nu_{6,10}-\nu_{6,11}+\nu_{6,{\bar 2}}+\nu_{6,{\bar 4}})^{q\lambda}_p,\nonumber\\
\hskip-10mm&&t_{p,sc}^{q\lambda}= 2(\nu_{6,7}+\nu_{6,{\bar
1}})_p^{q\lambda}),\quad t_{p,cc}^{q\lambda}=(\nu_{7,{\bar
1}})_p^{q\lambda}\,. \label{t_xi,xi'-expl}
\end{eqnarray}
Here the lower index $p$ and the upper indices $q\lambda$ at
brackets mean that they should be put at each term within brackets,
while the notation $\nu_{i,j}$ (used for clarity) means the same as
$\nu_{ij}$ in (\ref{nu_i,ij-def}). Quantities
$t_{\xi'\xi}^{q\lambda}$ with $\xi'\neq\xi$ not presented in Eqs.
(\ref{t_xi,xi'-expl}) can be obtained from those given in
(\ref{t_xi,xi'-expl}) by interchanging indices $q$ and $\lambda$:
$t_{\xi'\xi}^{q\lambda}=t_{\xi\xi'}^{\lambda q}$.

The above-discussed relations of symmetry can also be used to
calculate quantities $l_{p,n}^{\lambda}$, ${m}_{p,n}^q$ and
$t_{p,nm}^{q\lambda}$ in Eqs. (\ref{l_p,n})-(\ref{m,t_p^q,n}) used
in the  SSJA. This is illustrated below.

\subsection{Kinetic mean-field calculations\label{KMFA-calc}}

In this section we describe calculations of averages
$\langle\hat{w}_{01}^p\rangle$, $l_{p,n}^{\lambda}$, ${m}_{p,n}^q$
and $t_{p,nm}^{q\lambda}$  in  Eqs. (\ref{w_01^p}), (\ref{l_p,n}),
 (\ref{m,t_p^q,n}) using the simplest approximation
which neglects fluctuations of occupation numbers $n_i^p$: each
$n^p_i$ is replaced by its mean value $\langle n_i^p\rangle=c_p$. At
the same time, thermodynamic quantities, in particular, chemical
potentials $\lambda_{\rho}$ in Eqs. (\ref{Gamma_p}), will be found
using the more exact, pair-cluster approximation -- PCA (or
``cluster variation method for pair clusters'' \cite{VS-99}). It can
significantly raise the accuracy of calculations with respect to the
usual mean-field approximation (MFA), particularly  for dilute
alloys \cite{VKh-08,VZhKh-10}. To differ this our approach from the
usual MFA, we call it ``the kinetic mean-field approximation''
(KMFA).

Let us first find the KMFA expression for the average
$\overline{w}_p$  of the operator $\hat{w}_{01}^p$ given by Eq.
(\ref{w_01^p}). Replacing each $n^p_i$ in (\ref{w_01^p}) by the site
fraction $c_p$, we obtain:
\begin{equation}
\Big(\overline{w}_p\Big)_{\rm KMFA}\equiv
w_p^0=\Gamma_pc_h^2S_{p\Delta}^4S_u^{14}\,.\label{w_p^0}
\end{equation}
The upper index ``0'' at averages $w_p$, $\nu_p$, $m_p$, $l_p$ and
$t_p$ will mean ``KMFA'', and we denote for brevity:
\begin{equation}
S_{p\Delta}=\Big(1+\sum_{\lambda}c_{\lambda}f_{p\Delta}^{\lambda}\Big),\qquad
S_u=\Big(1+\sum_{\mu}c_{\mu}f_{u}^{\mu}\Big).\label{S_Delta,u}
\end{equation}
The factor  $\Gamma_p$ in (\ref{w_p^0}), according to
(\ref{Gamma_p}), can be expressed via the activation frequency
$\gamma_p$ and the chemical potentials $\lambda_{\rho}$ of vacancies
or solute atoms with respect to host atoms. Each $\lambda_{\rho}$ is
the sum of the ideal solution term $\lambda_{\rho}^{id}=T\ln
(c_{\rho}/c_h)$ and the interaction term $\lambda_{\rho}^{int}$:
\begin{eqnarray}
&&\beta\lambda_v=\ln\,(c_v/c_h)+\beta\lambda_v^{int}
\nonumber\\
&&\beta\lambda_{\alpha}=\ln\,(c_{\alpha}/c_h)
+\beta\lambda_{\alpha}^{int}\label{lambda_v,alpha}
\end{eqnarray}
In a dilute alloy, the interaction term $\lambda_{\rho}^{int}$ is
linear in the solute site fractions $c_{\alpha}$. We will describe
this term by the PCA expression which for dilute alloys becomes
exact \cite{VZhKh-10}. For a binary alloy, these expressions are
given below by Eqs. (\ref{a_v,B}), while for a multi-component
dilute alloy they can be obtained from Eqs. (26)-(31) in
\cite{VZhKh-10}:
\begin{eqnarray}
&&\beta\lambda_v^{int}=-\sum_{\gamma}\sum_{n=1}
z_nf_n^{v\gamma}c_{\gamma}\nonumber\\
&&\beta\lambda_{\alpha}^{int}=-\sum_{\gamma}\sum_{n=1}
z_nf_n^{\alpha\gamma}c_{\gamma}.\label{lambda^int-dilute}
\end{eqnarray}
Here $z_n$ is the coordination number for the $n$-th shell in the
crystal, and $f_n^{\rho\gamma}$ is the Mayer function for the
configurational interaction $v_n^{\rho\gamma}$ (\ref{v_ij-def}) in
this shell:
\begin{equation}
f_n^{\rho\gamma}=[\exp\,(\beta\,v_n^{\rho\gamma})-1]\,.\label{f_n^rho-gamma}
\end{equation}

Using Eqs. (\ref{Gamma_p}) and (\ref{lambda_v,alpha}), we can write
$w_p^0$ in (\ref{w_p^0}) as
\begin{equation}
w_p^0=c_p\omega_p \label{w_p-omega_p}
\end{equation}
where the quantity $\omega_p$ is defined as follows:
\begin{eqnarray}
&& \omega_{\alpha}=\gamma_{\alpha v}c_va_va_{\alpha}
S_{\alpha\Delta}^4S_u^{14}\nonumber\\
&& \omega_h=\gamma_{hv}c_va_v
S_{h\Delta}^4S_u^{14}\,.\label{omega-alpha,h}
\end{eqnarray}
Here the factor $a_v$ or $a_{\alpha}$ defined by the relation
\begin{equation}
a_v=\exp\,(\beta\lambda_v^{int}),\qquad  a_{\alpha}
=\exp\,(\beta\lambda_{\alpha}^{int})\label{a_v,a_alpha}
\end{equation}
can be called ``the reduced activity coefficient'' for a vacancy or
for a solute atom (our $a_{\alpha}$ is related to the activity
coefficient $\gamma_{\alpha}$ used, e. g., in \cite{Allnatt-93} as:
$a_{\alpha}$=$c_h\gamma_{\alpha}$).

When $c_{\alpha}\to 0$, factors $a_v$, $a_{\alpha}$, $S_{p\Delta}$,
$S_u$ in (\ref{S_Delta,u}) and (\ref{a_v,a_alpha}) tend to unity,
thus quantities $\omega_p$  in (\ref{omega-alpha,h}) take the values
\begin{equation}
\omega_{\alpha}^0=c_v\gamma_{\alpha v},\qquad
\omega_{h}^0=c_v\gamma_{hv}.\label{omega_alpha,h-0}
\end{equation}
Hence $\omega_{p}^0$  has the meaning of the mean frequency of the
vacancy-($p$-species atom) exchanges in a dilute alloy. For a
concentrated alloy, $\omega_p$ can be viewed as the average value of
this frequency found in the KMFA. Note that the mean frequency
$\omega_{\alpha}^0$ differs from the ``solute jump frequency''
$w_{\alpha}$ used in the standard five-frequency model
\cite{Lidiard-55,Lidiard-60,Manning-64,Howard-Man-67,Bocquet-74,LeClaire-78,Ishioka-84,Allnatt-93,
Bocquet-96} which is related to $\omega_{\alpha}^0$ as follows:
\begin{equation}
w_{\alpha}=\omega_{\alpha}^0e_1^{v\alpha},\qquad
e_1^{v\alpha}=\exp\,(\beta v_1^{v\alpha})\label{w-omega_alpha}
\end{equation}
where $v_1^{v\alpha}$ is the nearest-neighbor vacancy-solute
interaction. Factor  $e_1^{v\alpha}$ in (\ref{w-omega_alpha})
corresponds to the factor $\exp\,(\beta\hat{E}^{in}_{pi,vj})$ in
(\ref{W_ij^pv}), and it is canceled in the mean frequency
$\omega_{\alpha}^0$ due to the statistical averaging in Eqs.
(\ref{dc^alpha,h-dt}).

Discussing calculations of one-site averages $\nu_{p\xi}^q$ in Eqs.
(\ref{l_p-expl-3}), (\ref{m_p^q-l_p^q}) and (\ref{t_1p}) we first
note that the differences between  averages which include operators
of occupation of sites of the symmetry $v$, $s$ or $c$ mentioned
above arise only due to the inter-site correlations. Hence these
differences are not manifested in the KMFA. Therefore, each of
indices $v$, $s$, $c$ in Eqs. (\ref{l_p-expl-4})-(\ref{l_pv^lambda})
can be replaced by the common index $u$. Second, using identities
(\ref{identities-for-nu_i,ij}) we see that the average
$\nu_{p\xi}^q=\langle\hat{w}_{01}^pn^q_{\xi}\rangle$ differs from
the average $\langle\hat{w}_{01}^p\rangle=w_p^0$ by replacing one of
factors $S_{p\xi}$ in Eq. (\ref{w_p^0}) (with $S_{pu}\equiv S_u$) by
the appropriate factor $c_qe^q_{p\xi}$ with $e^q_{p\xi}$ from Eqs.
(\ref{e-exp}). It yields the following relations:
\begin{equation}
\nu_{p\Delta}^{q0}=c_pc_q\omega_p\eta_{p\Delta}^q,\qquad
\nu_{pu}^{q0}=c_pc_q\omega_p\eta_{u}^q \label{nu_p^q-KMFA}
\end{equation}
where we denote for brevity:
\begin{equation}
\eta_{p\Delta}^q=e^q_{p\Delta}/S_{p\Delta},\qquad
\eta_{u}^q=e^q_u/S_u.\label{eta_Delta,u}
\end{equation}
The same methods can be used for the KMFA calculations of two-site
averages $\nu_{p,ij}^{q\lambda}$ in  (\ref{t_xi,xi'-expl}). Hence
the KMFA expressions for one-site and two-site averages are similar:
\begin{equation}
\nu_{pi}^{q0}=c_pc_q\omega_p\eta_{p\xi}^q,\qquad
\nu_{p,ij}^{q\lambda,0}=c_pc_qc_{\lambda}\omega_p
\eta_{p\xi}^q\eta_{p\xi'}^{\lambda}\,. \label{nu_p^q-KMFA}
\end{equation}
Here indices $\xi$ and  $\xi'$ equal to $\Delta$ or $u$ indicate the
above-mentioned symmetry of site $i$ and site $j$, respectively, and
relations\, \,$\eta_{pu}^q\equiv\eta_{u}^q$,\,
\,$\eta_{pu}^{\lambda}\equiv \eta_{u}^{\lambda}$\, are implied.

The resulting KMFA expressions for quantities $m_{p,1}^q$,
$t_{1p}^{q\lambda}$ and $t_{2p}^{q\lambda}$ in Eqs.
(\ref{m_p^q-l_p^q}) and (\ref{t-t_1,2}) can be written as follows
\begin{eqnarray}
\hskip-5mm&m_{p,1}^{q0}=&c_pc_q\omega_p(2\eta_{p\Delta}^{q}-3\eta_{u}^{q})\nonumber\,,\\
\hskip-5mm&t_{1p}^{q\lambda,0}=&\delta_{q\lambda}c_pc_{\lambda}
\omega_p(2\eta_{p\Delta}^{q}+7\eta_{u}^{q})\nonumber\,,\\
\hskip-5mm&t_{2p}^{q\lambda,0}=&c_pc_qc_{\lambda}\omega_p
\Big[6\eta_{p\Delta}^{q}\eta_{p\Delta}^{\lambda}\nonumber\\
\hskip-5mm&&- 12(\eta_{p\Delta}^{q}\eta_{u}^{\lambda}+
\eta_{p\Delta}^{\lambda}\eta_{u}^{q})+11\eta_{u}^{q}\eta_{u}^{\lambda}\Big].
\label{KMFA-results-m,t}
\end{eqnarray}

Calculations of averages $l_{p,n}^{\lambda}$, $m_{p,n}^q$ and
$t_{p,nm}^{\,q\lambda}$ in Eqs. (\ref{l_p,n}),  (\ref{m,t_p^q,n})
for values $n,m>1$  corresponding to the SSJA can be made similarly
to those for the NNJA described above, though description of
rotations of vectors ${\bf R}_{l_n^+}$ and ${\bf R}_{l_n^-}$ in
(\ref{m,t_p^q,n}) (analogous to that given in Table I for vectors
${\bf R}_{1k}$) should be made for each $n$ and $m$ separately. The
results can be written in terms of ``reduced'' quantities
$\tilde{l}_{p,n}^{\lambda}$ and $\tilde{m}_{p,n}^q$ defined by the
relations:
\begin{eqnarray}
&&l_{p,n}^{\lambda
0}=c_{\lambda}c_p\omega_p\,\tilde{l}_{p,n}^{\lambda},\qquad
m_{p,n}^{q0}=c_qc_p\omega_p\,\tilde{m}_{p,n}^q\label{tilde-l,m}
\end{eqnarray}
where $\omega_p$ is the same as in Eqs. (\ref{omega-alpha,h}).
Expressions for quantities $\tilde{l}_{p,n}^{\lambda}$ and
$\tilde{m}_{p,n}^q$ in (\ref{tilde-l,m})  via $\eta_{p\Delta}^q$ and
$\eta_{u}^q$ in  (\ref{eta_Delta,u}) and the factor
\begin{equation}
\xi^q_u=(\eta_{u}^q-1)\label{nu^q_u}
\end{equation}
are given in Table II.

\begin{center}

\vbox{\noindent TABLE II. Reduced values $\tilde{l}_{p,n}^{\lambda}$
and $\tilde{m}_{p,n}^q$ in Eqs. (\ref{tilde-l,m})
 \vskip3mm
\begin{tabular}{|c|ccccc|}
\hline
&&&&&\\
$n$&1&2&3&4&5\\
\hline
&&&&&\\
$\tilde{l}_{p,n}^{\lambda}$&$(4\eta_{p\Delta}^{\lambda}-6\eta_{u}^{\lambda})$
&\hskip3mm$2\xi^{\lambda}_u$&\hskip3mm$4\xi^{\lambda}_u$
&\hskip3mm$4\xi^{\lambda}_u$&\hskip3mm$2\xi^{\lambda}_u$\\
&&&&&\\
$\tilde{m}_{p,n}^q$&$(2\eta_{p\Delta}^q-3\eta_{u}^q)$
&\hskip3mm$4\xi^q_u$&\hskip3mm$\xi^q_u$&\hskip3mm$2\xi^q_u$&\hskip3mm$\xi^q_u$\\
\hline
\end{tabular}}
\end{center}
\vskip3mm

\noindent Similarly, matrices $t_{p,nm}^{\,q\lambda}$ which enter
into Eqs. (\ref{m,t_p^q,n}) can be expressed via the ``reduced''
matrices $\tilde{t}_{p,nm}^{\lambda}$ and
$\tilde{t}_{p,nm}^{\,q\lambda}$:
\begin{equation}
t_{p,nm}^{\,q\lambda,0}=c_qc_p\omega_p\,\Big(\delta_{\,q\lambda}\tilde{t}_{1p,nm}^{\lambda}
+c_{\lambda}\tilde{t}_{2p,nm}^{\,q\lambda}\Big).\label{tilde-t}
\end{equation}
Here the matrix $\tilde{t}_{1p,nm}^{\lambda}$ has a relatively
simple form:
\begin{eqnarray} \hspace{-8mm}&&\left(\begin{array}{ccccc}
2\eta_{p\Delta}^{\lambda}+7\eta_{u}^{\lambda}&-\eta_{u}^{\lambda}&
-2\eta_{u}^{\lambda}&-2\eta_{u}^{\lambda}&-\eta_{u}^{\lambda}\\
&&&&\\
-4\eta_{u}^{\lambda}&4\eta_{u}^{\lambda}+8&0&-4&0\\
&&&&\\
-\eta_{u}^{\lambda}&0&2\eta_{u}^{\lambda}+9&-1&-1\\
&&&&\\
-2\eta_{u}^{\lambda}&-1&-2&2\eta_{u}^{\lambda}+10&-2\\
&&&&\\
-\eta_{u}^{\lambda}&0&-2&-2&\eta_{u}^{\lambda}+11\\
\end{array}\right)\nonumber
\end{eqnarray}
while the matrix $\tilde{t}_{2p,nm}^{\,q\lambda}$ can be written as
follows:
\begin{eqnarray} \hspace{-8mm}&&
\left(\begin{array}{ccccc} \tilde{t}_{2p,11}^{\,q\lambda}&
\chi_p^{q\lambda}&2\chi_p^{q\lambda}&2\chi_p^{q\lambda}&\chi_p^{q\lambda}\\
&&&&\\ 4\chi_p^{\lambda q}&
\tilde{t}_{2,22}^{\,q\lambda}&8\varepsilon^{q\lambda}&
8\varepsilon^{q\lambda}+4&4\varepsilon^{q\lambda}\\
&&&&\\
\chi_p^{\lambda q}&\varepsilon^{q\lambda}&
\tilde{t}_{2,33}^{\,\lambda\,q}&
2\varepsilon^{q\lambda}+1&\varepsilon^{q\lambda}+1\\
&&&&\\
2\chi_p^{\lambda q}&2\varepsilon^{q\lambda}+1
&4\varepsilon^{q\lambda}+2&
\tilde{t}_{2,44}^{\,q\lambda}&2\varepsilon^{q\lambda}+2\\
&&&&\\
\chi_p^{\lambda q}&\varepsilon^{q\lambda}
&2\varepsilon^{q\lambda}+2&
2\varepsilon^{q\lambda}+2&\tilde{t}_{2,55}^{\,q\lambda}\\
\end{array}\right)\nonumber
\end{eqnarray}
where the diagonal elements $\tilde{t}_{2p,nn}^{\,q\lambda}$ are:
\begin{eqnarray}
&&\tilde{t}_{2p,11}^{\,q\lambda}=6\eta_{p\Delta}^q\eta_{p\Delta}^{\lambda}
-12\Big(\eta_{p\Delta}^q\eta_{u}^{\lambda}+\eta_{u}^q\eta_{p\Delta}^{\lambda}\Big)
+11\eta_{u}^q\eta_{u}^{\lambda}\nonumber\\
&&\tilde{t}_{2,22}^{\,q\lambda}=4\Big(\eta_{u}^q\eta_{u}^{\lambda}-2\eta_{u}^q
-2\eta_{u}^{\lambda}\Big)\nonumber\\
&&\tilde{t}_{2,33}^{\,q\lambda}=\Big(2\eta_{u}^q\eta_{u}^{\lambda}-4\eta_{u}^q
-4\eta_{u}^{\lambda}-5\Big)\nonumber\\
&&\tilde{t}_{2,44}^{\,q\lambda}=2\Big(3\eta_{u}^q\eta_{u}^{\lambda}-4\eta_{u}^q
-4\eta_{u}^{\lambda}-1\Big)\nonumber\\
&&\tilde{t}_{2,55}^{\,q\lambda}=\Big(\eta_{u}^q\eta_{u}^{\lambda}-2\eta_{u}^q
-2\eta_{u}^{\lambda}-9\Big),\label{t_p,2-diagonal}
\end{eqnarray}
while non-diagonal elements are expressed via only two quantities,
$\chi_p^{q\lambda}$ and $\varepsilon^{q\lambda}$:
\begin{eqnarray}
&&\chi_p^{q\lambda}= \Big(4\eta_{p\Delta}^q\eta_u^{\lambda}
-5\eta_{u}^q\eta_{u}^{\lambda}-4\eta_{p\Delta}^q+6\eta_{u}^q\Big)\nonumber\\
&&\varepsilon^{q\lambda}=2\xi^q_u\xi^{\lambda}_u=2(\eta_{u}^q-1)(\eta_{u}^{\lambda}-1).
\label{t_p,2-nondiagonal}
\end{eqnarray}

The KMFA calculations described above neglect contributions of
fluctuations of occupation numbers $n_i^{\alpha}$ in the statistical
averages. Calculations of such contributions can be made using the
more refined statistical methods, such as the pair-cluster
approximation - PCA \cite{VS-99}, and they will be described
elsewhere together with their contribution to the enhancement of
chemical diffusion. At the same time, these calculations show that
these fluctuative contributions are typically not very significant,
and the above-described KMFA expressions are usually sufficient for
the realistic description of diffusion, particularly in dilute
alloys.

\section{DIFFUSION IN BINARY ALLOYS\label{Diff-binary-gen}}

\subsection{General expressions for Onsager coefficients in a binary alloy\label{Onsag-binary-gen}}

For a binary alloy $AB$ with $h=A$ and $\alpha=B$, fields
$h_n^{\alpha\alpha}$ in Eqs. (\ref{J^alpha,h-l_n}) and
(\ref{m-t-expl}) are zero due to the antisymmetry property
(\ref{h_ij-minus_h_ji}), thus Eqs. (\ref{m-t-expl})  take the form
of a system of  $n_{max}$ equations  for $n_{max}$ fields
$h_n^{\alpha v}$=$h_n^{Bv}$:
\begin{eqnarray}
&&\sum_{m=1}^{n_{max}}a_{nm} h_m^{B
v}=(m_{B,n}^A\delta\mu_{Bv}-{m}_{A,n}^{B}\delta\mu_{Av}),\nonumber\\
&& a_{nm}= \Big({t}_{h,nm}^{BB}-t_{B,nm}^{AB}
-2m_{B,n}^h\delta_{m1}\Big)  \label{A_nm}
\end{eqnarray}
where $\delta_{m1}$ is unity when $m=1$ and zero otherwise.

In the NNJA, Eqs. (\ref{A_nm}) include only one field $h_1^{\alpha
v}$ which is simply expressed via  $m_{p,1}^q$ and
$t_{p,11}^{q\lambda}$ in (\ref{m_p^q-expl})-(\ref{t-t_1,2}):
\begin{eqnarray}
&&h_1^{Bv}=(m_{B,1}^A\delta\mu_{Bv}-
m_{A,1}^{B}\delta\mu_{Av})/a_{11}\,,\nonumber\\
&&a_{11}=\Big(t_{A,11}^{BB}-t_{B,11}^{AB}- 2m_{B,1}^A\Big).
\label{h_1-nn-jump}
\end{eqnarray}
Substituting this $h_1^{Bv}$ in Eq. (\ref {J^alpha,h-l_n}) with
$n_{max}=1$ and using also Eq. (\ref{m_p^q-l_p^q}), we obtain the
following relations between fluxes $J^{p}_{0\to 1}$ and  differences
$\delta\mu_{qv}$:
\begin{eqnarray}
& J^{B}_{0\to 1}=&-\beta\delta\mu_{Bv} \Big[\overline{w}_{B}+
2m_{B,1}^A(\overline{w}_{B}+m_{B,1}^{B})/a_{11}\Big]\nonumber\\
&& +\beta
\delta\mu_{Av}2m_{A,1}^{B}(\overline{w}_{B}+m_{B,1}^{B})/a_{11}
\nonumber\\
& J^{A}_{0\to 1}=& -\beta \delta\mu_{B
v}2m_{A,1}^{B}m_{B,1}^A/a_{11}\nonumber\\
&&- \beta \delta\mu_{Av}\Big[\overline{w}_{A}
-2(m_{A,1}^{B})^2/a_{11}\Big] \label{J^p-gen}
\end{eqnarray}
which determine the Onsager coefficients $L_{pq}$ in (\ref{L-A_pq}).

Note that the Onsager symmetry relation,
\begin{equation}
L_{BA}=L_{AB}, \label{L_sym}
\end{equation}
in our approach is obeyed identically. According to Eq.
(\ref{J^p-gen}), Eq. (\ref{L_sym}) implies:
\begin{equation}
m_{B,1}^{B}+m_{B,1}^A=-\overline{w}_{B}. \label{m-sym}
\end{equation}
Using Eqs. (\ref{l_p-expl-3})-(\ref{m_p^q-l_p^q}) we can re-write
(\ref{m-sym}) as
\begin{equation}
\langle\hat{w}_{B}[2(n_2^{B}+n_2^A)-2(n_6^{B}+n_6^A)-
(n_7^{B}+n_7^A)]\rangle=-\langle\hat{w}_{B}\rangle \label{w-m-sym}
\end{equation}
which holds identically as $(n_i^{B}+n_i^A)\equiv 1$.

One can show that the symmetry relation (\ref{L_sym}) holds also for
the SSJA. Presence of this relation irrespectively of site fractions
and approximations used illustrates the theoretical consistency of
the  master equation approach.

Using Eqs. (\ref{J^p-gen}) and (\ref{L_sym}), we can write the
general NNJA expressions for Onsager coefficients $L_{pq}$ as
follows:
\begin{eqnarray}
&& (T/na_0^2)L_{BB}=\Big[w^0_B
-2(m_{B,1}^h)^2/a_{11}\Big]\,,\nonumber\\
&&(T/na_0^2)L_{BA}=L_{BA}=
\Big(2m_{A,1}^{B} m_{B,1}^A/a_{11}\Big)\,,\nonumber\\
&& (T/na_0^2)L_{AA}=\Big[w^0_A-2(m_{A,1}^{B})^2/a_{11}\Big].\,
\label{L_pq-bin-nn-jump}
\end{eqnarray}

To write explicit expressions for $L_{pq}$  in
(\ref{L_pq-bin-nn-jump}), it is convenient to omit index $\alpha=B$
of the only kind of solute atoms in the site fraction $c_{\alpha}$
and in quantities $\eta_{p\Delta}^{\alpha}$, $e_{p\Delta}^{\alpha}$,
$\eta^{\alpha}_u$ and $e^{\alpha}_u$ defined by  Eqs.
(\ref{eta_Delta,u}), and also to employ the ``reduced'' denominator
$D_{nn}$ rather than the quantity $a_{11}$ from (\ref{h_1-nn-jump}),
as well as the frequency ratio $z =\omega_B/\omega_A$ rather than
the frequency $\omega_B$ from (\ref{omega-alpha,h}):
\begin{eqnarray}
&& c_B=c,\quad \eta_{A\Delta}^{B}=\eta_{A\Delta},\quad
e_{A\Delta}^{B}=e_{A\Delta},\nonumber\\
&& \eta^{B}_u=\eta_u,\quad e^{B}_u=e_u\,,\quad
a_{11}=cc_A\omega_AD_{nn},\nonumber\\
&& z=\omega_B/\omega_A=\gamma_{Bv}a_{B} S_{B\Delta}^4/ \gamma_{Av}
S_{A\Delta}^4\,.\label{D_nn,z,c_B=c}
\end{eqnarray}
Using KMFA expressions (\ref{omega-alpha,h}), (\ref{tilde-l,m}),
(\ref{tilde-t}) for quantities $w_p^0$, $m_{h,1}^{\alpha}$,
$m_{\alpha,1}^h$ and $t_{p,11}^{q\lambda}$ in (\ref{h_1-nn-jump})
and (\ref{L_pq-bin-nn-jump}), we can write the NNJA  expressions for
Onsager coefficients as follows:
\begin{eqnarray}
&(T/na_0^2)L_{AA}=&
\omega_Ac_A\Big[1-2c(3\eta_{u}-2\eta_{A\Delta})^2/D_{nn}\Big],\nonumber\\
&(T/na_0^2)L_{AB}=&\omega_Bcc_A2(3\eta_{u}-2\eta_{A\Delta})\nonumber\\
&&\times
(3\eta_{u}^A-2\eta_{B\Delta}^A)/D_{nn}\,,\nonumber\\
&(T/na_0^2)L_{BB}=&
\omega_Bc\Big[1-2c_Az\nonumber\\
&&\times(3\eta_{u}^A-2\eta_{B\Delta}^A)^2/D_{nn}\Big].
\label{L_pq-nn-jump}
\end{eqnarray}
The denominator $D_{nn}$ in (\ref{L_pq-nn-jump}) can be conveniently
written as the sum of two terms, that without the common factor of
site fraction $c$ and that which includes this factor:
\begin{equation}
D_{nn}=(d_{1,11}+c\,d_{2,11}).\label{D_nn}
\end{equation}
Here quantities $d_{1,11}$ and $d_{2,11}$ are expressed via the
reduced parameters $\tilde{m}_{p,1}^{q}$,
$\tilde{t}_{1p,11}^{\lambda}$, and $\tilde{t}_{2p,11}^{q\lambda}$ in
(\ref{tilde-l,m}) and (\ref{tilde-t}) in accordance with Eqs.
(\ref{h_1-nn-jump}) and (\ref{D_nn,z,c_B=c}):
\begin{eqnarray}
&&d_{1,11}=(\tilde{t}_{1A,11}^{B}-2\tilde{m}_{B,1}^{A}),\nonumber\\
&& d_{2,11} =(\tilde{t}_{2A,11}^{BB}-
z\,\tilde{t}_{2B,11}^{AB})\label{A_1,2-11}
\end{eqnarray}
or, explicitly:
\begin{eqnarray}
\hskip-10mm&&d_{1,11}=(2\eta_{A\Delta}+7\eta_{u})+2z(3\eta_{u}^A
-2\eta_{B\Delta}^A),\label{A_1-11-expl}\\
\hskip-10mm&&d_{2,11}=\Big(6\eta_{A\Delta}^2-24\eta_{A\Delta}\eta_u
+11\eta_u^2\Big)\nonumber\\
\hskip-10mm&&-z\Big[6\eta_{B\Delta}^A\eta_{B\Delta}
-12(\eta_{B\Delta}^A\eta_u+\eta_{B\Delta}\eta_u^A)
+11\eta_u^A\eta_u\Big]. \label{A_2-11-expl}
\end{eqnarray}
In Eqs. (\ref{D_nn,z,c_B=c})-(\ref{A_1,2-11}), quantities
$\eta_{B\Delta}^A$,  $\eta_u^A$,  $\eta_{A\Delta}$ and $\eta_u$ are
defined by Eqs. (\ref{eta_Delta,u}) and (\ref{e-h-1}):
\begin{eqnarray}
&&\eta_{B\Delta}^A=1/S_{B\Delta},\quad
\eta_u^A=1/S_u\,,\quad\eta_{A\Delta}=e_{A\Delta}/S_{A\Delta},\nonumber\\
&&\eta_u=e_u/S_u,\qquad \eta_{B\Delta}=e_{B\Delta}/S_{B\Delta},
\nonumber\\
&&S_{A\Delta}=(1+cf_{A\Delta}),\qquad S_{B\Delta}=(1+cf_{B\Delta}),
\nonumber\\
&&S_{u}=(1+cf_u),\qquad  f_{A\Delta}=(e_{A\Delta}-1), \nonumber\\
&&f_{B\Delta}=(e_{B\Delta}-1),\qquad f_u=(e_u-1),\label{eta-expl}
\end{eqnarray}
while factors $e_{A\Delta}$ and $e_u$ in (\ref{eta-expl}) are
defined by Eqs. (\ref{D_nn,z,c_B=c}),  (\ref{e-exp}) and
(\ref{f_Delta,u}).

Let us also explicitly write the reduced activity coefficients $a_v$
and $a_{B}$ in (\ref{a_v,a_alpha}). Using for chemical potentials
$\lambda_v$ and $\lambda_B$ their PCA expressions given by Eqs. (39)
in Ref. \cite{BV-98}, we obtain for coefficients $a_v$ and $a_{B}$
in (\ref{a_v,a_alpha}):
\begin{eqnarray}
\hskip-10mm&&a_v=\exp\,\Big\{-\sum_{n=1}z_n\ln[1+2cf_n^{vB}/(R_n+1)]\Big\}\,, \nonumber\\
\hskip-10mm&&a_B=\exp\,\Big\{-\sum_{n=1}z_n\ln[1+2cf_n^{BB}/(R_n+1)]\Big\}.
\label{a_v,B}
\end{eqnarray}
Here $z_n$, $f_n^{vB}$  and $f_n^{BB}$ are the same as in
(\ref{lambda^int-dilute}):
\begin{equation}
f_n^{vB}=\exp (-\beta v_n^{vB})-1\,,\quad f_n^{BB}=\exp (-\beta
v_n^{BB})-1\,, \label{f_n^{BB},v}
\end{equation}
while $R_n$ is expressed via $f_n^{BB}$ as follows:
\begin{equation}
R_n=(1+4cc_Af_n^{BB})^{1/2}.\label{R_n}
\end{equation}

For the SSJA, the general expressions for Onsager coefficients in a
concentrated binary alloy can be obtained from Eqs.
(\ref{tilde-l,m})-(\ref{t_p,2-diagonal}) similarly to the NNJA
expressions (\ref{L_pq-nn-jump}). However, these  general
expressions are cumbersome. Therefore, these expressions will be
given elsewhere in connection with their contributions to the
enhancement of chemical diffusion, while in Sec. \ref{dilute
-binary} we present them only for the case of a dilute alloy.

\subsection{Expressions for chemical diffusion coefficients and
correlation factors in a concentrated binary
alloy\label{D_p-f_p-concent-alloys}}

In this section we present explicit expressions for the chemical (or
``intrinsic'' \cite{Allnatt-93}) diffusion coefficients $D_A$ and
$D_B$. First we discuss the thermodynamic ``activity factor''
$A_{ac}$ which enters into these expressions. As vacancies for
processes under consideration are locally equilibrium and their
chemical potential $\mu_v$ is zero
 \cite{Allnatt-93}, differences $\mu_{qv}=(\mu_{q}-\mu_{v})$ in Eq.
(\ref{J-nabla-mu}) can be replaced by absolute chemical potentials
$\mu_q=\partial F/\partial N_q$ where $F$ is the total free energy
and $N_q$ is the total number of $q$-species atoms. These $\mu_q$
are related to quantity $\lambda_{\alpha}=\lambda_B$ in Eqs.
(\ref{dc^alpha,h/dt-lin}) and the grand canonical potential per
atom, $\Omega$, by the following relations \cite{VZhKh-10}:
\begin{equation}
\mu_A=\Omega,\qquad \mu_B=\lambda_B+\Omega\, \label{mu_A,B}
\end{equation}
where the PCA expression for $\Omega$ is presented in
\cite{VZhKh-10}:
\begin{eqnarray}
\hskip-10mm&&\Omega=T\ln c_A\nonumber\\
\hskip-10mm&&-{1\over 2}
T\sum_{n=1}z_n\ln\left[1-2c^2f_n^{BB}/(R_n+1+2cf_n^{BB})\right].
\label{Omega_PCA}
\end{eqnarray}
while $z_n$ and $f_n^{BB}$ are the same as in (\ref{a_v,B}). The
diffusion coefficients $D_p$ are defined by the Fick's first law
\cite{Allnatt-93}:
\begin{equation}
{\bf J}_A=-D_A\nabla n_A\,,\qquad {\bf J}_B=-D_B\nabla n_B\,
\label{D_A,B-def}
\end{equation}
where ${\bf J}_p$ is the same as in Eq. (\ref{J-nabla-mu}) and $n_p$
is the density of $p$-species atoms. To write explicit expressions
for $D_p$, we can also use the Gibbs-Duhem relation:
\begin{equation}
c_Ad\mu_A+c_Bd\mu_B=c_A\,d\Omega+c\,d\mu_B=0 \label{Duhem}
\end{equation}
which for Eqs. (\ref{mu_A,B}), (\ref{Omega_PCA}) and the PCA
expression for $\lambda_B$ given by Eqs. (\ref{lambda_v,alpha}),
(\ref{a_v,a_alpha}) and (\ref{a_v,B}) can also be checked by a
direct calculation. Using Eqs. (\ref{J-nabla-mu}), (\ref{D_A,B-def})
and (\ref{Duhem}) and supposing that the mean volume $\bar v$ per
atom of an alloy is described by  the Vegard's law:
\begin{equation}
\bar v=\overline{v}_Ac_A+\overline{v}_Bc_B \label{Vegard}
\end{equation}
where $\overline{v}_p$ is the atomic volume of a $p$-component in an
alloy, we can write $D_p$ as follows:
\begin{eqnarray}
&& D_A=(T/n^2\overline{v}_A)\Big(L_{AA}/c_A-L_{AB}/c\Big)A_{ac}\,,
\nonumber\\
&& D_B=(T/n^2\overline{v}_B)\Big(L_{BB}/c-L_{AB}/c_A\Big)A_{ac}\,.
\label{D_A,B}
\end{eqnarray}
Here Onsager coefficients $L_{pq}$ are the same as in
(\ref{L_pq-nn-jump}), while the activity factor $A_{ac}$  is related
to the reduced activity $a_B$ in Eq. (\ref{a_v,a_alpha}) by the
following relation:
\begin{equation}
A_{ac}=1+cc_Ad\ln a_B/dc=1+cc_Ad(\beta \lambda_B^{int})/dc\,.
\label{A_ac}
\end{equation}
Substituting $\lambda_B^{int}=T\ln a_B$ with $a_B$ from
(\ref{a_v,B}), we obtain:
\begin{eqnarray}
\hskip-10mm&&A_{ac}=1\nonumber\\
\hskip-10mm&&-cc_A\sum_{n=1}z_n2f_n^{BB}{\big[R_n+1-4c(1-2c)f_n^{BB}\big]\over
(R_n+1)(R_n+1+2cf_n^{BB})}.\label{A_ac-PCA}
\end{eqnarray}

Let us now discuss the Onsager coefficients  $L_{pq}$ in
(\ref{D_A,B}). Eqs. (\ref{L_pq-nn-jump}) show that each $L_{pq}$ can
be conveniently expressed via the mean frequency $\omega_p$ and the
reduced ``correlative'' coefficients $L_{pq}^c$ which describe
vacancy correlation effects and are defined by the following
relations:
\begin{eqnarray}
&&(T/na_0^2)L_{AA}= \omega_Ac_A(1-cL^c_{AA}),
\nonumber\\
&&(T/na_0^2)L_{AB}=\omega_Bcc_AL^c_{AB}\,,\nonumber\\
&&(T/na_0^2)L_{BB}= \omega_Bc(1-c_AL^c_{BB}). \label{L_pq^c-def}
\end{eqnarray}
Using  Eqs. (\ref{L_pq-nn-jump}),  we can concisely write the
correlative coefficients $L_{pq}^c$ in  (\ref{L_pq^c-def}) for the
NNJA as follows:
\begin{eqnarray}
&&L_{AA}^c= 2(3\eta_u-2\eta_{A\Delta})^2/D_{nn}\,,
\nonumber\\
&&L_{AB}^c=2(3\eta_u-2\eta_{A\Delta})(3\eta_u^A-2\eta_{B\Delta}^A)/D_{nn}\,,\nonumber\\
&&L_{BB}^c=2z(3\eta_u^A-2\eta_{A\Delta}^A)^2/D_{nn}\,.
\label{L_pq^c-nn}
\end{eqnarray}
Eqs. (\ref{D_A,B}) and (\ref{L_pq^c-def}) show that each diffusion
coefficient is proportional to several factors of different nature:
the mean frequency $\omega_p$, the correlation factor $f_p$, and the
activity factor $A_{ac}$, similarly to the dilute alloy case
\cite{Allnatt-93}:
\begin{eqnarray}
&&D_p=(a_0^2/n\overline{v}_p)\omega_pf_pA_{ac}\nonumber\\
&&f_A=1-(zc_AL^c_{AB}+cL^c_{AA}),\nonumber\\
&&f_B=1-(c_AL^c_{BB}+cL^c_{AB})\label{D_p-f_p}
\end{eqnarray}
but factors $\omega_p$, $f_p$, $A_{ac}$ depend on the solute
fraction $c$. Applications of Eqs. (\ref{D_p-f_p})  to the
description of enhancement  of chemical diffusion will be described
elsewhere.

\subsection{Onsager coefficients in a dilute binary
alloy\label{dilute -binary}}

In the dilute alloy limit $c\to 0$, frequencies $\omega_p$ tend to
$\omega_p^0$ in (\ref{omega_alpha,h-0}), while parameters $z$,
$\eta_{B\Delta}^A$, $\eta_{A\Delta}$, $\eta^A_u$ and $\eta_u$ in
(\ref{L_pq-nn-jump})-(\ref{A_2-11-expl}), according to
(\ref{D_nn,z,c_B=c}) and (\ref{eta-expl}), take the following
values:
\begin{eqnarray}
&&z_0=\gamma_{Bv}/\gamma_{Av},\,\quad(\eta_{B\Delta}^A)_0=(\eta^A_u)_0=1,\nonumber\\
&&(\eta_{A\Delta})_0=e_{A\Delta},\quad \eta_u^0=e_u.
\label{z_0-eta_0}
\end{eqnarray}
Here and below, the upper or the lower index ``0'' at each quantity
indicates its value at $c=0$.

To relate our notation to that commonly used for the five-frequency
model
\cite{Lidiard-55,Lidiard-60,Manning-64,Howard-Man-67,Bocquet-74,LeClaire-78,Ishioka-84,Allnatt-93,
Bocquet-96} we note that the jump rates (``frequencies'') $w_n$ of
that model in our notation are:
\begin{eqnarray}
&&w_0=\omega_A^0,\quad w_1=\omega_A^0e_{A\Delta}e_1^{vB},\quad
w_2=\omega_B^0e_1^{vB},\nonumber\\
&&w_3=\omega_A^0e_1^{vB}e_u,\quad w_4=\omega_A^0e_u \label{w_n-e_i}
\end{eqnarray}
where $e_1^{vB}$ is the same as in (\ref{w-omega_alpha}). At the
same time, exponential factors $e_u$ and $e_{A\Delta}$ in
(\ref{w_n-e_i}) have a more clear physical meaning than frequencies
$w_n$. Eqs. (\ref{D_nn})-(\ref{eta-expl}) include also the factor
$e_{B\Delta}$ analogous to $e_{A\Delta}$ which describes influence
of a solute atom $B$ near the bond $(ij)$ on the
$Bi\leftrightharpoons vj$ jump probability. Thus below we use
instead of $w_n$ quantities $x_n$ and $y_1$ defined as follows:
\begin{eqnarray}
&&x_1=e_{A\Delta},\qquad x_2=\omega_B^0/\omega_A^0,\qquad
x_4=e_u,\nonumber\\
&& y_1=e_{B\Delta}=\exp[\beta (2u_1^B-\Delta_{B}^B)]\label{x_n-def}
\end{eqnarray}
with $x_2$ equal to $z_0$ in (\ref{z_0-eta_0}).   In this notation,
the Mayer functions $f_{p\Delta}$ and $f_u$ and the low-$c$ values
of factors $\xi_u$ and $\xi_u^A$ in  Eqs. (\ref{eta-expl}) take the
following form:
\begin{eqnarray}
&&f_{A\Delta}=(x_1-1),\quad f_{B\Delta}=(y_1-1),\nonumber\\
&&f_u=(x_4-1),\quad \xi_u=f_u=(x_4-1),\nonumber\\
&&\xi_u^A(c\ll 1)=(-cf_u)=-c(x_4-1).\label{f-x,y}
\end{eqnarray}

Below  we present the low-$c$ expansions for mean frequencies
$\omega_p$ and  Onsager coefficients $L_{pq}$ up to the first order
in $c$, and the zero-order terms for correlative Onsager
coefficients $L_{pq}^c$ and correlation factors $f_p$ in Eqs.
(\ref{L_pq^c-def})-(\ref{D_p-f_p}). The fluctuative corrections
mentioned in Sec. \ref{KMFA-calc} make no contribution to these
terms, hence we can use the KMFA expressions (\ref{L_pq^c-nn}).

Let us first consider the mean frequency $\omega_p$ and define the
enhancement factor $b_p^{\omega}$ for it by the usual relation:
\begin{equation}
\omega_p(c)=\omega_p^0(1+c\,b_p^{\omega}).\label{b_p^omega}
\end{equation}
Using  Eqs. (\ref{omega-alpha,h})  for $\omega_p$ and the PCA
expressions (\ref{a_v,B}) for activity factors $a_v$ and $a_B$, we
find:
\begin{eqnarray}
\hskip-10mm&&b_A^{\omega}=(4f_{A\Delta}+14f_u+b_{vB})\label{b_A^omega}\,
\nonumber\\
\hskip-10mm&&b_B^{\omega}=(4f_{B\Delta}+14f_u+b_{vB}+b_{BB})\,.
\label{b_A,B^omega}
\end{eqnarray}
Here $f_{p\Delta}$ and $f_u$ are the same as in (\ref{f-x,y}), while
$b_{vB}$ and $b_{BB}$ are contributions of the activity factors,
$a_v$ and $a_B$:
\begin{eqnarray}
\hskip-8mm&&b_{vB}=-\sum_{n=1}z_nf_n^{vB}=-12f_1^{vB}-6f_2^{vB}-\ldots\label{b_v}\\
\hskip-8mm&&b_{BB}=-\sum_{n=1}z_nf_n^{BB}=-12f_1^{BB}-6f_2^{BB}-\ldots\,.
\label{b_B}
\end{eqnarray}
For correlative terms and correlation factors at $c=0$,
$L_{pq}^{c0}$ and  $f_{p0}$, we find from Eqs.
(\ref{L_pq^c-nn})-(\ref{f-x,y}) in the NNJA:
\begin{eqnarray}
\hskip-3mm&&L_{AA}^{c0}=2(3x_4-2x_1)^2/D_0\,,\quad
L_{AB}^{c0}=2(3x_4-2x_1)/D_0\,,\nonumber\\
\hskip-3mm&&L_{BB}^{c0}=2x_2/D_0\,,\qquad D_0=(2x_1+2x_2+7x_4)\,.
\label{L-pq_0-nn}\\
\hskip-3mm&&f_{A0}=1-x_2L_{AB}^{c0},\qquad f_{B0}=1-L_{BB}^{c0}\,.
\label{f_p0}
\end{eqnarray}
%

In the SSJA, values of $L_{pq}^{c0}$ can be obtained from the
general expressions for  $L_{pq}$ mentioned in Sec.
\ref{Onsag-binary-gen}:
\begin{eqnarray}
&L_{AA}^{c0}=&2\Big[(3x_4-2x_1)^2-2(3x_4-2x_1)\rho_0f_u\nonumber\\
&&+\rho_0D_0f_u^2/x_4\Big]/D_{ss}^0\,,\nonumber\\
&L_{AB,0}^c=&2(3x_4-2x_1-\rho_0f_u)/D_{ss}^0\,,\nonumber\\
&L_{BB,0}^c=&2x_2/D_{ss}^0\,,\qquad D_{ss}^0=D_0-\rho_0x_4\,,
\label{L^c0-ss}
\end{eqnarray}
while correlation factors $f_{p0}$ are expressed via these
$L_{pq}^{c0}$ according to Eqs. (\ref{f_p0}). $D_0$ in
(\ref{L^c0-ss}) is the same as in (\ref{L-pq_0-nn}), while $\rho_0$
is related to the ``vacancy escape function'' $F=F(x_4)$ of the
five-frequency model \cite{Allnatt-93} as:
\begin{eqnarray}
&&\rho_0=7(1-F)=P_N(x_4)/P_D(x_4)\label{rho^0_i}
\end{eqnarray}
where  $P_N(x)$ and $P_D(x)$ are polynomials found by Bocquet
\cite{Bocquet-74}:
\begin{eqnarray}
\hskip-10mm&&P_N(x)=(10x^4+190x^3+1031x^2+1594.5x),\nonumber\\
\hskip-10mm&&P_D(x)=(2x^4+45x^3+328x^2+930.5x+855.5).\,
\label{F,p_N,D}
\end{eqnarray}
For a more accurate description of vacancy correlation effects at
low $c$ (used, in particular, in Sec. \ref{estimates-of-x_n}), the
polynomials $P_N$ and $P_D$ in (\ref{rho^0_i}) can be taken from
Ref. \cite{Manning-64}.

Using Eqs. (\ref{L_pq^c-def}) and (\ref{b_p^omega}), we can write
Onsager coefficients at low $c$ as follows:
\begin{eqnarray}
&&(T/na_0^2)L_{AA}=\omega_A^0\,[1+c(b_A^{\omega}-1-L_{AA}^{c0})]
\nonumber\\
&&(T/na_0^2)L_{AB}=\omega_B^0\,cL_{AB}^{c0}. \nonumber\\
&&(T/na_0^2)L_{BB}=\omega_B^0\,c\,(1-L_{BB}^{c0}).
\label{L_pq-dilute}
\end{eqnarray}
For the case of very low vacancy concentration under consideration:
$c_v\ll c_B$, values of  $L_{pq}$ in Eqs. (\ref{L_pq-dilute}) with
\,$b_A^{\omega}$\, and \,$L_{pq,0}^c$\, given by Eqs.
(\ref{b_A,B^omega})-(\ref{L^c0-ss}) coincide with those found in the
traditional theory \cite{Allnatt-93}.  At the same time, the
expression for $L_{AA}$ obtained by Nastar \cite{Nastar-05}
corresponds to missing the vacancy-solute interaction term $b_{vB}$
given by Eq. (\ref{b_A,B^omega})  in the term $b_A^{\omega}$ in
(\ref{L_pq-dilute}).

\section{ENHANCEMENT OF TRACER SOLVENT DIFFUSION IN A DILUTE
BINARY ALLOY \label{b_A^*-sec}}

Below we discuss enhancement of diffusion of radioactive isotopes
(``tracers'') in a dilute alloy $AB$. The tracer solvent enhancement
factor $b_{A^*}$ was calculated in a number of previous studies
reviewed in detail by Nastar \cite{Nastar-05}. However, some
significant contribution to $b_{A^*}$ discussed below was missed in
these studies. The tracer solute enhancement factor $b_{B^*}$, to
our knowledge, was not calculated, even though experimental values
of this factor are known for a number of alloys \cite{LeClaire-78}.

For simplicity,  below we use the simplest approximations for both
the statistical calculations and the description of vacancy
correlation effects employing KMFA and NNJA. Going beyond these
approximations, as well as a possible influence of non-pairwise
effective interactions $h_{ijk}^{\rho\sigma\tau}$ in
(\ref{h_rho-sigma}) (discussed by Barbe and Nastar \cite{Barbe-06}
for other problems), will be discussed  elsewhere in calculations of
enhancement factors for chemical diffusion. At the same time,
comparison of our results with the available Monte Carlo simulations
seems to imply that the effects disregarded in this work make
usually a relatively weak influence on the tracer enhancement
factors in real alloys.

Therefore, our calculations of both $b_{A^*}$ and $b_{B^*}$  use the
NNJA equations (\ref{m-t-expl})  for fields $h_1^{\rho\sigma}$ and
the KMFA expressions (\ref{KMFA-results-m,t}) for coefficients in
these equations.

\subsection{General equations of diffusion in a ternary alloy
\label{b_p^*-subsec}}

For a ternary alloy which contains components $\alpha$, $\beta$ and
$h$, Eqs. (\ref{m-t-expl})  take the following form:
\begin{eqnarray}
\hskip-5mm&&(m_{\alpha}^h\delta\mu_{\alpha}-{m}_{h}^{\alpha}\delta\mu_{h})+h_{\alpha
v}(2m_{\alpha}^h-t_{1h}^{\alpha}+
t_{2\alpha}^{h\alpha}-t_{2h}^{\alpha\alpha})
\nonumber\\
&&\hskip10mm+h_{\beta v}(t_{2\alpha}^{h\beta}-t_{2h}^{\alpha\beta})
+h_{\alpha\beta}t_{2\alpha}^{h\beta}=0\,,
\nonumber\\
\hskip-5mm&&(m_{\beta}^h\delta\mu_{\beta}-{m}_{h}^{\beta}\delta\mu_h)+h_{\alpha
v}(t_{2\alpha}^{h\alpha}-t_{2h}^{\beta\alpha})
\nonumber\\
&&\hskip10mm +h_{\beta v}(2m_{\beta}^h-t_{1h}^{\beta}+
t_{2\beta}^{h\beta}-t_{2h}^{\beta\beta})
-h_{\alpha\beta}t_{2\beta}^{h\alpha}=0\,,
\nonumber\\
\hskip-5mm&&(m_{\alpha}^{\beta}\delta\mu_{\alpha}-{m}_{\beta}^{\alpha}\delta\mu_{\beta})+h_{\alpha
v}(2m_{\alpha}^{\beta}-t_{1\beta}^{\alpha}+
t_{2\alpha}^{\beta\alpha}-t_{2\beta}^{\alpha\alpha})
\nonumber\\
&&\hskip10mm+h_{\beta v}(2m_{\beta}^{\alpha}-t_{1\alpha}^{\beta}+
t_{2\beta}^{\alpha\beta}-t_{2\alpha}^{\beta\beta})\nonumber\\
&&\hskip10mm+h_{\alpha\beta}(t_{1\alpha}^{\beta}+t_{1\beta}^{\alpha}+
t_{2\alpha}^{\beta\beta}+t_{2\beta}^{\alpha\alpha})=0\,.
\label{ternary-Eq-gen}
\end{eqnarray}
Here we replace each difference $\delta\mu_{pv}$ by $\delta\mu_{p}$,
as  in Sec. \ref{D_p-f_p-concent-alloys}; omit the common index
$n$=$m$=1 at $h_n^{\rho\sigma}$ and $m_{p,n}^q$:
$h_{\rho\sigma}$=$h_1^{\rho\sigma}$,\, $m_{p}^q$=$m_{p,1}^q$;\, and
use Eq. (\ref{t-t_1,2}) to express $t_{p,11}^{q\lambda}$ via
one-site and two-site averages, $t_{1p}^{\lambda}$ and
$t_{2p}^{q\lambda}$.

Writing Eqs. (\ref{J^alpha,h-l_n}) for fluxes $J^p_{0\to 1}$ in a
ternary alloy, it is convenient to use the identity
\begin{equation}
m_p^{\alpha}+m_p^{\beta}+m_p^h=-\overline{w}_p
\label{Onsag-rel-tern}
\end{equation}
which is the evident generalization of Eq. (\ref{m-sym}) for a
binary alloy. Then Eqs. (\ref{J^alpha,h-l_n}) can be written as
follows:
\begin{eqnarray}
\hskip-5mm&&-T J^{\alpha}_{0\to 1}={w}_{\alpha}^0 \delta\mu_{\alpha}
-2m_{\alpha}^hh_{\alpha v}+2m_{\alpha}^{\beta}(h_{\beta
v}+h_{\alpha\beta}-h_{\alpha v}),
\nonumber\\
\hskip-5mm&&-T J^{\beta}_{0\to 1}={w}_{\beta}^0 \delta\mu_{\beta}
-2m_{\beta}^hh_{\beta v}+2m_{\beta}^{\alpha}(h_{\alpha
v}-h_{\alpha\beta}-h_{\beta v}),
\nonumber\\
\hskip-5mm&&-T J^h_{0\to 1}=e{w}_{h}^0
\delta\mu_{h}+2m_{\alpha}^hh_{\alpha v}+2m_{\beta}^hh_{\beta v}
\label{J^p-gen-tern}
\end{eqnarray}
where we also take into account relations (\ref{h_ij-minus_h_ji})
and (\ref{m_p^q-l_p^q}).

In Eqs. (\ref{ternary-Eq-gen})-(\ref{J^p-gen-tern}), the tracer
self-diffusion corresponds to  $h=A$,  $\beta=B$, $\alpha=A^*$, and
the tracer solute diffusion, to $h=A$, $\beta=B$, $\alpha=B^*$. Each
tracer diffusion coefficient $D_{p^*}$ with $p^*$ equal to $A^*$ or
$B^*$ can be written in the general form (\ref{D_p-f_p}) with
replacing $p$ by $p^*$ and the correlation factor $f_{p^*}$ which
corresponds tothe terms with fields $h_{p\rho}$ in Eqs.
(\ref{J^p-gen-tern}). The enhancement factor $b_{p^*}$ is defined by
the usual relation:
\begin{equation}
D_{p^*}(c)=D_{p^*}^0(1+c\,b_{p^*}) \label{b_p^*}
\end{equation}
and, as in Eq. (\ref{D_p-f_p}), it includes three different terms:
\begin{eqnarray}
\hskip-10mm&&b_{p^*}=b_{p^*}^{\omega}+b_{p^*}^c\,+b_{p^*}^{ac}.
\label{b_p^*-sum}
\end{eqnarray}
Here the frequency enhancement factor $b^{\omega}_{p^*}$ was
discussed in Sec. \ref{dilute -binary}, while the correlation
enhancement factor $b^c_{p^*}$ and the activity enhancement factor
$b_{p^*}^{ac}$ (related to the activity  $A_{p^*}^{ac}$) are defined
similarly to $b_{p^*}$ in (\ref{b_p^*}):
\begin{equation}
b_{p^*}^c={\partial\over\partial c}\Big[\ln
b_{p^*}(c)\Big]_0\,,\qquad b_{p^*}^{ac}={\partial\over\partial
c}\Big[A_{p^*}^{ac}(c)\Big]_0 \label{b_p_*^c,ac}
\end{equation}
where the lower index ``0'' means the $c\to 0$ value of the
derivative, and we take into account that $A_{p^*}^{ac}(0)$ is 1.

\subsection{Calculation of enhancement factor for tracer
self-diffusion in a dilute binary alloy \label{tracer-b_A^*}}

For the tracer self-diffusion, we should put in Eqs.
(\ref{ternary-Eq-gen}) $h=A$,  $\beta=B$, $\delta\mu_{\beta}=0$, and
$\alpha=A^*$, but to make formulas compact, we will also employ
symbol $\alpha$ instead of $A^*$. We consider the case of a low site
fraction of solute: $c\ll 1$, and the tracer site fraction
$c_{\alpha}$ in the real experiments is low, too. However, Eqs.
(\ref{ternary-Eq-gen}) can be easily solved for any $c_{\alpha}$
which enables us to discuss also some methodical problems. Thus at
first we consider the case of the arbitrary $c_{\alpha}$.

Let us discuss different terms in (\ref{b_p^*-sum}). The frequency
enhancement factor  $b_{A^*}^{\omega}$ is defined by Eqs.
(\ref{b_p^omega}) and  (\ref{omega-alpha,h}) and, as atoms $A^*$ and
$A$ are chemically identical, it coincides with that for chemical
diffusion given by the first equation (\ref{b_A,B^omega}). The
activity enhancement factor $b_{A^*}^{ac}$ can be found from Eqs.
(\ref{A_ac}) and (\ref{A_ac-PCA}) with replacing $B$ by $A^*$, which
implies: $A^{ac}_{A^*}=1$, $b_{A^*}^{ac}=0$ Thus only the
correlation enhancement factor $b_{A^*}^{c}$ should by calculated.

Expressing averages $m_p^q$, $t_{1p}^q$ and $t_{2p}^{qr}$ in
(\ref{ternary-Eq-gen}) via reduced quantities $\tilde{m}_p^q$,
$\tilde{t}_{1p}^q$ and $\tilde{t}_{2p}^{qr}$ defined in
(\ref{tilde-l,m})-(\ref{t_p,2-diagonal}), and replacing index
$\alpha$ in these quantities by index $A$ due to the chemical
identity of atoms $A^*$ and $A$, we can write Eqs.
(\ref{ternary-Eq-gen}) for the tracer self-diffusion at low $c$ as
follows:
\begin{eqnarray}
\hskip-5mm&&\tilde{m}_A^A(\delta\mu_{\alpha}-\delta\mu_{A})
+h_{\alpha v}(2\tilde{m}_A^A-\tilde{t}_{1A}^{A})\nonumber\\
&&\hskip5mm +ch_{\alpha B}\tilde{t}_{2A}^{AB}=0,
\nonumber\\
\hskip-5mm&&-\tilde{m}_{A}^B\delta\mu_{A}+h_{\alpha v}c_{\alpha}
(x_2\tilde{t}_{2B}^{AA}-\tilde{t}_{2A}^{BA})-h_{\alpha
B}c_{\alpha}x_2\tilde{t}_{2B}^{AA}
\nonumber\\
&&\hskip5mm +h_{Bv}(2x_2\tilde{m}_B^A-\tilde{t}_{1A}^{B})=0\,,
\nonumber\\
\hskip-5mm&&\tilde{m}_{A}^B\delta\mu_{\alpha} +h_{\alpha
v}(2\tilde{m}_{A}^B-x_2\tilde{t}_{1B}^{A}+
c_{\alpha}\tilde{t}_{2A}^{BA}-c_{\alpha}x_2\tilde{t}_{2B}^{AA})
\nonumber\\
\hskip-5mm&&\hskip5mm +h_{\alpha
B}c_{\alpha}(\tilde{t}_{1A}^B+x_2\tilde{t}_{1B}^A
+c_{\alpha}x_2\tilde{t}_{2B}^{AA})\nonumber\\
\hskip-5mm&&\hskip5mm
-h_{Bv}(2x_2\tilde{m}_B^A-\tilde{t}_{1A}^{B})=0\,,
\label{Eq-h-A^*-1}
\end{eqnarray}
while Eqs.  (\ref{J^p-gen-tern}) for fluxes take the form:
\begin{eqnarray}
\hskip-5mm&&-T J^{\alpha}_{0\to 1}={\omega}_A[c_{\alpha}
\delta\mu_{\alpha} -2c_{\alpha}c_A\tilde{m}_A^Ah_{\alpha
v}\nonumber\\
\hskip-5mm&&\hskip15mm+ 2c\,c_{\alpha}\tilde{m}_A^B(h_{Bv}
+h_{\alpha B}-h_{\alpha v})],
\nonumber\\
\hskip-5mm&&-T J^A_{0\to 1}={\omega}_A[c_A\delta\mu_A
+2c_{\alpha}c_A\tilde{m}_A^Ah_{\alpha
v}+2c\,c_A\tilde{m}_A^Bh_{Bv}],
\nonumber\\
\hskip-5mm&&-T J^B_{0\to
1}=2c\,{\omega}_B\tilde{m}_B^A[c_{\alpha}(h_{\alpha
v}-h_{\alpha B}-h_{Bv})-c_Ah_{Bv}].\nonumber\\
\hskip-5mm \label{J^p-A^*}
\end{eqnarray}
Explicit expressions for coefficients  $\tilde{m}_p^q$,
$\tilde{t}_{1p}^q$ and $\tilde{t}_{2p}^{qr}$ needed to solve Eqs.
(\ref{Eq-h-A^*-1}) and (\ref{J^p-A^*}) at low $c$ can be found using
Eqs. (\ref{KMFA-results-m,t}), (\ref{eta-expl}), (\ref{z_0-eta_0})
and (\ref{x_n-def}):
\begin{eqnarray}
\hskip-5mm&&\tilde{m}_A^A=-1+c(3f_u-2f_{A\Delta}),\quad
\tilde{t}_{1A}^{A}=9-c(7f_u+2f_{A\Delta}),\nonumber\\
\hskip-5mm&&
\tilde{t}_{2A}^{AB}=-(6x_1+x_4),\quad\tilde{m}_{A}^B=(2x_1-3x_4),\quad
\tilde{t}_{2B}^{AA}=-7,\nonumber\\
\hskip-5mm&& \tilde{m}_B^A=-1,\quad
\tilde{t}_{1B}^{A}=9,\quad\tilde{t}_{1A}^B=(2x_1+7x_4).
\label{tilde-m,t-expl}
\end{eqnarray}
Taking into account the Gibbs-Duhem relation (\ref{Duhem}) for $A$
and $A^*$ atoms:
\begin{equation}
c_{\alpha}\delta\mu_{\alpha}+c_A\delta\mu_A=0 \label{Gibbs-Duhem-2}
\end{equation}
and the relation $c_{\alpha}$+$c_A$=$(1-c)\simeq$1, we can express
$\delta\mu_{\alpha}$ and $\delta\mu_A$ in (\ref{Eq-h-A^*-1}) via the
difference $\delta\mu=\delta\mu_{\alpha}-\delta\mu_A$:
\begin{equation}
\delta\mu_{\alpha}=c_A\delta\mu,\qquad
\delta\mu_A=-c_{\alpha}\delta\mu\,. \label{delta-mu-alpha,A}
\end{equation}
Then sum of two last equations and the first equation in the system
(\ref{Eq-h-A^*-1}) yield two equations for $h_{\alpha v}$ and
$h_{\alpha B}$:
\begin{eqnarray}
\hskip-5mm&&\tilde{m}_{A}^B\delta\mu +h_{\alpha
v}(2\tilde{m}_{A}^B-x_2\tilde{t}_{1B}^{A}) +h_{\alpha
B}(\tilde{t}_{1A}^B+x_2\tilde{t}_{1B}^A)=0,\nonumber\\
\hskip-5mm&&\tilde{m}_A^A\delta\mu +h_{\alpha
v}(2\tilde{m}_A^A-\tilde{t}_{1A}^{A})+ch_{\alpha
B}\tilde{t}_{2A}^{AB}=0\,.\label{Eq-h-A^*-2}
\end{eqnarray}

Solving Eqs. (\ref{Eq-h-A^*-1}) and (\ref{Eq-h-A^*-2}) up to the
first order in $c$, we can write each field $h_{p\rho}$ as
$(h_{p\rho}^0+ch_{p\rho}^1)$. Contributions of these fields to
fluxes $J^p_{0\to 1}$ in (\ref{J^p-A^*})  include terms $h_{\alpha
v}^0$, $h_{\alpha B}^0$, $h_{B v}^0$ and $h_{\alpha v}^1$ given by
such expressions:
\begin{eqnarray}
\hskip-5mm&&h_{\alpha v}^0=-\delta\mu/11,\quad
h_{Bv}^0=c_{\alpha}(h_{\alpha v}^0-h_{\alpha B}^0), \nonumber\\
\hskip-5mm&&h_{\alpha B}^0=-\delta\mu\,
9(2x_1+x_2-3x_4)/11(2x_1+9x_2+7x_4),\nonumber\\
\hskip-5mm&&h_{\alpha v}^1=[20(x_1-x_4)h_{\alpha
v}^0-(6x_1+x_4)h_{\alpha B}^0]/11. \label{h_0,1-expl}
\end{eqnarray}
Now we note that for the tracer self-diffusion with any tracer site
fraction $c_{\alpha}$, both the total flux of $A$ and  $A^*$ atoms
and the flux of $B$ atoms should be absent \cite{Allnatt-93}:
\begin{equation}
J^{A}_{0\to 1}+J^{\alpha}_{0\to 1}=0, \qquad J^{B}_{0\to 1}=0\,.
\label{J^A+A^*,B=0}
\end{equation}
Using Eqs.  (\ref{J^p-A^*}) we see that  both relations
(\ref{J^A+A^*,B=0}) hold true for the solutions (\ref{h_0,1-expl}).
It illustrates the theoretical consistency of these solutions.

Going to the physical results, we consider  the realistically low
tracer site fractions $c_{\alpha}\ll 1$.  Substituting solutions
(\ref{h_0,1-expl}) into the first equation (\ref{J^p-A^*}), using
Eq. (\ref{J_p-J^p_01}) to relate quantity $J^{\alpha}_{0\to 1}$ to
the flux $J^{\alpha}$, and Eq. (\ref{D_A,B-def}) (with replacing
$B\to A^*$), to relate this flux to the tracer diffusion coefficient
$D_{A^*}$, we obtain:
\begin{equation}
D_{A^*}=a_0^2\omega_{A^*}f_{A^*}^0(1+c\,b_{A^*}^c).
\label{D_A^*-expl}
\end{equation}
Here $f_{A^*}^0=9/11$ is the well-known value of the tracer
self-diffusion correlation factor in the NNJA used
\cite{Allnatt-93}, while the correlation enhancement factor
$b_{A^*}^c$ is:
\begin{eqnarray}
\hskip-5mm&&b_{A^*}^c=-{2\over 9}-{4\over 11}\left[
(2x_1-3x_4)+{2(8x_1-17x_4)^2\over 9(2x_1+9x_2+7x_4)}\right].\nonumber\\
\hskip-5mm\label{b_A^*c}
\end{eqnarray}
The frequency enhancement factor $b_{A^*}^{\omega}$ was mentioned to
coincide with $b_{A}^{\omega}$ in (\ref{b_A,B^omega}):
\begin{equation}
b_{A^*}^{\omega}=(4x_1+14x_4-18+b_{vB})\label{b_A^*-omega}
\end{equation}
with $b_{vB}$ given by  Eq. (\ref{b_v}), while the activity term
$b_{A^*}^{ac}$ in (\ref{b_p^*-sum}) is zero. Therefore, the total
self-diffusion enhancement factor $b_{A^*}$ (\ref{b_p^*}) has the
form
\begin{equation}
b_{A^*}=b_{A^*}^{\omega}+b_{A^*}^c\label{b_A^*-tot}
\end{equation}
with $b_{A^*}^{\omega}$ and $b_{A^*}^c$ given by Eqs.
(\ref{b_A^*-omega}) and (\ref{b_A^*c}).

Presence of vacancy-solute term $b_{vB}$ (\ref{b_v}) in Eqs.
(\ref{b_A^*-omega}) and  (\ref{b_A^*-tot}) implies, in particular,
that the vacancy-solute binding energies $(-v_1^{vB})$ can hardly be
as high as those found recently for some plausible models
\cite{Faupel-88,Hagenschulte-89}. For example, value
$(-v_1^{vB})\sim 0.26$ eV suggested in \cite{Hagenschulte-89} for
alloys AgSb at $T$ between 1048 and 890 K corresponds to $b_{vB}$
between (-200) and \hbox{(-350)} which can hardly be compatible with
the experimental $b_{A^*}$ between about 30 and 50, even if the
entropy terms discussed in \cite{Hagenschulte-89} (and disregarded
in the five-frequency model) are taken into account. Values
$(-v_1^{vB})$ in the interval 5-100 meV given in Table III below and
typical for theoretical estimates \cite{Hagenschulte-94} seem to be
more realistic.

\subsection{Discussion of previous calculations of $b_{A^*}$ \label{b_A^*-discussion}}

Previous calculations of $b_{A^*}$ using both the traditional
methods \cite{Lidiard-60,Howard-Man-67,Ishioka-84,Allnatt-93} and
the Monte Carlo simulations \cite{Belova-03a,Belova-03b} were
reviewed in detail by Nastar together with her original results
\cite{Nastar-05}. Let us first compare our results with those of
Nastar obtained using her version of the master equation approach.
There are two differences between our results in Eqs.
(\ref{b_A^*c})-(\ref{b_A^*-tot}) and those of Nastar given by Eqs.
(56) and (57) in \cite{Nastar-05}.

(i) The frequency enhancement factor  $b_{A^*}^{\omega}$ in
(\ref{b_A^*-omega}) includes the vacancy-solute interaction term
$b_{vB}$ which is absent in the analogous Eq. (56) in
\cite{Nastar-05}.

(ii) The constant term\, $(-2/9)$\, in the expression (\ref{b_A^*c})
for $b_{A^*}^c$ is absent in the analogous Eq. (57) in
\cite{Nastar-05}.

Disagreement (i) seems to be due to the general shortcoming of the
approach used by Nastar \cite{Nastar-05} mentioned in Secs.
\ref{Introduction},  \ref{Gen-eqs},  and \ref{dilute -binary} which
is related to the employing for finding of statistical averages of
some indirect considerations rather than the direct calculations.
Disagreement (ii) can be related just to a numerical error.

\begin{figure}
\caption{(color online) Dependencies of the tracer self-diffusion
enhancement factor $b_{A^*}$ on  $x_2$ at various $x_1$ and $x_4$ in
(\ref{x_n-def}) in the absence of vacancy-solute term $b_{vB}=0$ in
(\ref{b_A^*-omega}). Symbol MC correspond to Monte Carlo
simulations, and symbols L, HM, IK, N and PW, to Refs.
\cite{Lidiard-60}, \cite{Howard-Man-67}, \cite{Ishioka-84},
\cite{Nastar-05}, and the present work, respectively. Different
frames correspond to the following values of $x_1$ and $x_4$ and the
following MC simulations: (a) $x_1=2$, $x_4=1$, \cite{Belova-03b};\,
(b) $x_1=x_4=0.1$, \cite{Belova-03a};\, (c) $x_1=10x_2$, $x_4=0.1$,
\cite{Nastar-05};\, (d) $x_1=x_4=x_2$, \cite{Nastar-05}.\,  In
frames (c) and (d), curves N and PW merge within accuracy of
drawing. \label{b_A^*-x}}
\end{figure}

In Fig. 3 (where we use the data shown in Figs. 1-5 of Ref.
\cite{Nastar-05}), the results of various calculations of $b_{A^*}$
are compared with the available Monte Carlo simulations. Note that
the scale of parameters $x_n$ used in these simulations usually
differs from that typical for the real alloys for which $x_n$ and
$b_{A^*}$ usually obey the relations \cite{Allnatt-93,Bocquet-96}:
\begin{equation}
x_1,x_4\gtrsim 1,\quad x_2\gg 1, \quad |b_{A^*}|\gg
1\,.\label{x_n,b_A^*-exp}
\end{equation}
The realistic relations (\ref{x_n,b_A^*-exp}) are approximately
obeyed only for $x_n$ values used in frame \ref{b_A^*-x}$a$, and for
this frame, our calculations of $b_{A^*}$  agree with the Monte
Carlo simulations notably better than other calculations. For the
rest frames \ref{b_A^*-x}$b$\,-\,\ref{b_A^*-x}$d$, our calculations
also usually agree with the Monte Carlo simulations better than
other calculations,  but the relatively small scale of the
correlation term: $b_{A^*}^c\ll b_{A^*}^{\omega}$, makes differences
between different results to be less pronounced. Note also that
calculations of $b_{A^*}$ by Howard and Manning \cite{Howard-Man-67}
(used in the most of estimates of parameters of five-frequency model
for real alloys \cite{Allnatt-93,Bocquet-96}) usually differ from
Monte Carlo simulations stronger than those in \cite{Nastar-05} and
in the present work.

The comparison of relations (\ref{x_n,b_A^*-exp}) with
(\ref{b_A^*c})-(\ref{b_A^*-tot}) seems to imply that the main
contribution to $b_{A^*}$ is usually made by the frequency term
$b_{A^*}^{\omega}$, while the correlation term $b_{A^*}^c$ is less
important. It can be related to the presence in Eq.
(\ref{b_A^*-omega}) of large numerical factors both in the first two
terms, $(4f_{A\Delta}+14f_u)$, and in the third term, $b_{vB}\simeq
(-12f_{vB})$. Thus taking into account the vacancy-solute term
$b_{vB}$ missed in the previous calculations seems to be necessary
for realistic calculations of $b_{A^*}$.

\section{ENHANCEMENT OF TRACER SOLUTE DIFFUSION IN A DILUTE
BINARY ALLOY\label{b_B^*-sec}}

\subsection{General equations for tracer solute diffusion in a binary alloy
\label{b_B^*-gen-eq}}

Considering tracer solute diffusion, we put in Eqs.
(\ref{ternary-Eq-gen}) $h=A$,  $\alpha=B$, $\delta\mu_h=0$, and
$\beta=B^*$, but again we will also employ a compact symbol $\beta$
instead of $B^*$.

Expressing averages $m_p^q$, $t_{1p}^q$ and $t_{2p}^{qr}$ in
(\ref{ternary-Eq-gen}) via reduced quantities $\tilde{m}_p^q$,
$\tilde{t}_{1p}^q$ and $\tilde{t}_{2p}^{qr}$ defined in
(\ref{tilde-l,m})-(\ref{t_p,2-diagonal}), and replacing index
$\beta=B^*$ in these quantities and in $\omega_{\beta}$ in
(\ref{omega-alpha,h}) by index $B$ due to the chemical identity of
atoms $B^*$ and $B$, we can write Eqs. (\ref{ternary-Eq-gen}) for
tracer solute diffusion at any site fractions $c_{\beta}$ and $c_B$
as follows:
\begin{eqnarray}
\hskip-5mm&&z\tilde{m}_B^Ac_{\beta}\delta\mu_{\beta}
+c_{\beta}h_{\beta v}[(2z\tilde{m}_A^B-\tilde{t}_{1A}^{B})+
c_{\beta} (z\tilde{t}_{2B}^{AB}-\tilde{t}_{2A}^{BB})]
\nonumber\\
&&\hskip10mm
+c_{\beta}c_Bh_{Bv}(z\tilde{t}_{2B}^{AB}-\tilde{t}_{2A}^{BB})+
c_{\beta}c_Bh_{\beta B}z\tilde{t}_{2B}^{AB}=0,
\nonumber\\
\hskip-5mm&&z\tilde{m}_B^Ac_{B}\delta\mu_{B}+ c_{\beta}c_Bh_{\beta
v}(z\tilde{t}_{2B}^{AB}-\tilde{t}_{2A}^{BB})\nonumber\\
\hskip-5mm&&\hskip10mm+
c_Bh_{Bv}[(2z\tilde{m}_A^B-\tilde{t}_{1A}^{B})+
c_{\beta} (z\tilde{t}_{2B}^{AB}-\tilde{t}_{2A}^{BB})]\nonumber\\
\hskip-5mm&&\hskip10mm- c_{\beta}c_Bh_{\beta
B}z\tilde{t}_{2B}^{AB}=0,
\nonumber\\
\hskip-5mm&&\tilde{m}_B^Bc_{\beta}(\delta\mu_{\beta}-\delta\mu_B)
+c_{\beta}(h_{\beta v}-h_{Bv})[(2\tilde{m}_B^B-\tilde{t}_{1B}^{B})\nonumber\\
\hskip-5mm&&\hskip10mm+c_{\beta}h_{\beta
B}[2\tilde{t}_{1B}^{B}+(c_{\beta}+c_B)\tilde{t}_{2B}^{BB}]=0\,
\label{Eq-h-B^*-gen}
\end{eqnarray}
where $z=\omega_B/\omega_A$ is the same as in (\ref{D_nn,z,c_B=c}).
Atomic fluxes (\ref{J^p-gen-tern}) for tracer solute diffusion have
the following form:
\begin{eqnarray}
\hskip-5mm&&-T J^{\beta}_{0\to 1}={\omega}_B[c_{\beta}
\delta\mu_{\beta} -2c_{\beta}c_A\tilde{m}_B^Ah_{\beta
v}\nonumber\\
\hskip-5mm&&\hskip15mm+ 2c_{\beta}c_B\tilde{m}_B^B(h_{Bv}-h_{\beta
v} +h_{\beta B})],
\nonumber\\
\hskip-5mm&&-T J^B_{0\to 1}={\omega}_B[c_{B} \delta\mu_{B}
-2c_{B}c_A\tilde{m}_B^Ah_{B v}\nonumber\\
\hskip-5mm&&\hskip15mm+2c_{\beta}c_B\tilde{m}_B^B(h_{\beta
v}-h_{Bv}-h_{\beta B})],
\nonumber\\
\hskip-5mm&&-T J^A_{0\to
1}=2{\omega}_Ac_{A}\tilde{m}_A^B(c_{\beta}h_{\beta v}+c_{B}h_{B v}),
\hskip-5mm \label{J^p-B^*}
\end{eqnarray}
while coefficients  $\tilde{m}_p^q$, $\tilde{t}_{1p}^q$ and
$\tilde{t}_{2p}^{qr}$ in Eqs. (\ref{Eq-h-B^*-gen}) and
(\ref{J^p-B^*}) are defined by Eqs. (\ref{KMFA-results-m,t}),
(\ref{tilde-l,m}) and (\ref{tilde-t}).

Summing two first equations (\ref{Eq-h-B^*-gen}) and taking into
account the Gibbs-Duhem relation (\ref{Gibbs-Duhem-2}) for $B$ and
$B^*$ atoms:\, $c_{\beta}\delta\mu_{\beta}+c_B\delta\mu_B=0$, we
find that fields $h_{pv}$ in Eqs. (\ref{Eq-h-B^*-gen}) and
(\ref{J^p-B^*}) are related as follows:
\begin{equation}
c_{\beta}h_{\beta v}+c_Bh_{Bv}=0\,.\label{h-beta,B-v}
\end{equation}
Substituting this relation into Eqs. (\ref{J^p-B^*}) we see that
both the total flux of atoms $B$ and  $B^*$ and the flux of atoms
$A$ are absent at any site fractions $c_{\beta}$ and $c_B$:
\begin{equation}
J^{B}_{0\to 1}+J^{\beta}_{0\to 1}=0, \qquad J^{A}_{0\to 1}=0\,.
\label{J^B+B^*,A=0}
\end{equation}
Presence of these physically evident relations [analogous to
(\ref{J^A+A^*,B=0}) for tracer self-diffusion] illustrates
consistency of the theoretical approach used.

\subsection{Calculation of enhancement factor for tracer
solute diffusion in a dilute binary alloy \label{tracer-b_A^*}}

Below we calculate the tracer solute enhancement factor $b_{B^*}$ in
a dilute binary alloy $AB$ for the realistically low values of
tracer site fraction $c_{\beta}$:\,  $c_{\beta}\ll c$,\, $c=c_B\ll
1$. Using  Eq. (\ref{h-beta,B-v}) to express field $h_{Bv}$ via
$h_{\beta v}$, we can write Eqs. (\ref{Eq-h-B^*-gen}) for $h_{\beta
v}$ and $h_{\beta B}$ as follows:
\begin{eqnarray}
\hskip-5mm&&h_{\beta v}(\tilde{t}_{1A}^{B}-2z\tilde{m}_A^B)-
h_{\beta B}\,cz\tilde{t}_{2B}^{AB}=z\tilde{m}_B^A\delta\mu_{\beta},
\nonumber\\
\hskip-5mm&&h_{\beta v}(\tilde{t}_{1B}^{B}-2\tilde{m}_B^B)- h_{\beta
B}\,2\tilde{t}_{1B}^{B}=\tilde{m}_B^B\delta\mu_{\beta}
\label{Eq-h-B^*-2}
\end{eqnarray}
while Eq. (\ref{J^p-B^*}) for the tracer flux  ${\bf J}^{\beta}$
takes the form
\begin{eqnarray}
&&-T J^{\beta}_{0\to 1}={\omega}_Bc_{\beta} [\delta\mu_{\beta}
-2c_A\tilde{m}_B^Ah_{\beta v}\nonumber\\
&&\hskip22mm- 2c\tilde{m}_B^B(h_{\beta v}-h_{\beta B})].
\label{J^beta-B^*}
\end{eqnarray}
To explicitly write flux  ${\bf J}^{\beta}$ in (\ref{J^beta-B^*}),
we define the ``reduced'' fields $h_{\beta v}^0$, $h_{\beta B}^0$,
and $h_{\beta v}^1$ which correspond to solutions of Eqs.
(\ref{Eq-h-B^*-2}) in the zero-order and the first-order in $c$,
respectively, and express $h_{\beta v}$ via $h_{\beta v}^1$:
\begin{eqnarray}
\hskip-5mm&&h_{\beta
v}=\delta\mu_{\beta}(z\tilde{m}_{B}^A/D+ch_{\beta v}^1),\quad
h_{\beta v}^0=z_0\tilde{m}_{B,0}^A/D_0,
\nonumber\\
\hskip-5mm&&h_{\beta B}^0=[h_{\beta
v}^0(\tilde{t}_{1B,0}^{B}-2\tilde{m}_{B,0}^B)- \tilde{m}_{B,0}^B]
/(\tilde{t}_{1B,0}^{B}-2\tilde{m}_{B,0}^B),\nonumber\\
\hskip-5mm&& h_{\beta v}^1=h_{\beta
B}^0z_0\tilde{t}_{2B,0}^{AB}/D_0\,. \label{h^0,1-def}
\end{eqnarray}
where index ``0'' at each quantity indicates its value at $c=0$, as
in (\ref{z_0-eta_0}) and (\ref{L-pq_0-nn}), and $D$ [equal to
$d_{1,11}$ in (\ref{A_1,2-11})] is the coefficient at $h_{\beta v}$
in the first equation (\ref{Eq-h-B^*-2}):
\begin{equation}
D=(\tilde{t}_{1A}^{B}-2z\tilde{m}_B^A). \label{D-def}
\end{equation}
Using this notation and also Eqs. (\ref{J_p-J^p_01}) and
(\ref{D_A,B-def}), we can write  Eq. (\ref{J^beta-B^*})  in the form
of relation for the tracer solute diffusivity $D_{\beta}$ analogous
to Eq. (\ref{D_p-f_p}):
\begin{equation}
D_{B^*}=a_0^2\omega_{B^*}f_{B^*}A^{ac}_{B^*}.\label{D_beta-1}
\end{equation}
Due to the chemical identity of atoms $B^*$ and $B$, frequency
$\omega_{B^*}=\omega_{B}$ in (\ref{D_beta-1}) is given by Eq.
(\ref{omega-alpha,h}) for $\alpha=B$, the activity factor
$A^{ac}_{B^*}$ coincides with that in Eqs. (\ref{A_ac}) and
(\ref{A_ac-PCA}), and $f_{B^*}$ is the correlation factor:

\begin{eqnarray}
\hskip-10mm &&f_{B^*}=\Big[1-2zc_A(\tilde{m}_{B}^A)^2/D\nonumber\\
\hskip-10mm &&\hskip5mm +2c\Big(\tilde{m}_{B,0}^Bh_{\beta
B}^0-\tilde{m}_{B,0}^Bh_{\beta v}^0-\tilde{m}_{B,0}^Ah_{\beta
v}^1\Big)\Big].\label{f_B^*}
\end{eqnarray}
Explicit expressions for the low-$c$ values of functions $z$,
$\tilde{m}_p^q$, $\tilde{t}_{1p}^q$ and $\tilde{t}_{2p}^{qr}$ in
Eqs. (\ref{h^0,1-def})-(\ref{f_B^*}) can be found using Eqs.
(\ref{KMFA-results-m,t}), (\ref{D_nn,z,c_B=c}), (\ref{eta-expl}),
(\ref{z_0-eta_0}) and (\ref{x_n-def}):
\begin{eqnarray}
\hskip-5mm&&z=x_2[1+c(4f_{B\Delta}-4f_{A\Delta}+b_{BB})],\nonumber\\
\hskip-5mm&&
\tilde{m}_B^A=-1+c(3f_u-2f_{B\Delta}),\nonumber\\
\hskip-5mm&&\tilde{t}_{1A}^{B}=(2x_1+7x_4)-c(2x_1f_{A\Delta}+7x_4f_u),\nonumber\\
\hskip-5mm&&\tilde{m}_{B,0}^B=(2y_1-3x_4),\quad
\tilde{t}_{1B,0}^B=(2y_1+7x_4),\nonumber\\
\hskip-5mm&& \tilde{t}_{2B,0}^{AB}=-(6y_1+x_4)
\label{tilde-m,t-expl-2}
\end{eqnarray}
where Mayer functions $f_{A\Delta}$, $f_{B\Delta}$ and $f_u$ are the
same as in (\ref{eta-expl}), and  $b_{BB}$ is the same as in
(\ref{b_B}). Eqs. (\ref{f_B^*}),  \ref{tilde-m,t-expl-2}) and
(\ref{L-pq_0-nn}) show, in particular, that at $c\to0$, the
correlation factor for tracer solute diffusion is equal to that for
chemical diffusion \cite{Allnatt-93}:
\begin{equation}
f_{B^*0}=f_{B0}=(1-2x_2/D_0)\label{f_B0}
\end{equation}
where $D_0=(2x_1+2x_2+7x_4)$ is the same as in (\ref{L-pq_0-nn}).

According to Eqs. (\ref{b_p_*^c,ac}) and (\ref{f_B^*}), the
correlation term $b^c_{B^*}$ in (\ref{b_p^*-sum}) can be written as
the sum of two contributions, $b_{1B^*}^c$ and $b_{2B^*}^c$, which
correspond to the second and the third term in Eq. (\ref{f_B^*}):
\begin{eqnarray}
\hskip-10mm&&b^c_{B^*}=b_{1B^*}^c +b_{2B^*}^c\nonumber\\
\hskip-10mm&&b_{1B^*}^c=-{2\over f_{B0}}{\partial\over\partial
c}\Big[
zc_A(\tilde{m}_{B}^A)^2/D\Big]_0\nonumber\\
\hskip-10mm&&b_{2B^*}^c={2\over
f_{B0}}\Big(\tilde{m}_{B,0}^Bh_{\beta B}^0-\tilde{m}_{B,0}^Bh_{\beta
v}^0-\tilde{m}_{B,0}^Ah_{\beta v}^1\Big) \label{b_1,2B^*^c}
\end{eqnarray}
where $f_{B0}$ is given by Eq. (\ref{f_B0}).

Term $b_{1B^*}^c$ is the sum of three terms which correspond to
three factors in square brackets in (\ref{b_1,2B^*^c}):
\begin{equation}
b_{1B^*}^c=-(2x_2/f_{B0}D_0)(l_1+2l_2-l_3) \label{b_1B^*}
\end{equation}
where $l_n$ is the appropriate logarithmic derivative:
\begin{eqnarray}
\hskip-5mm&&l_1={\partial\over\partial
c}\ln (zc_A)\Big|_0=(4y_1-4x_1-1+b_{BB})\,,\nonumber\\
\hskip-5mm&&l_2={\partial\over\partial c}\ln
(\tilde{m}_{B}^A)\Big|_0=(2y_1-3x_4+1)\,,\nonumber\\
\hskip-5mm&&l_3={\partial\over\partial c}\ln
D\Big|_0=[2x_2(6y_1-4x_1-3x_4+b_{BB})\nonumber\\
\hskip-5mm&&\hskip25mm -2x_1^2-7x_4^2+D_0]/D_0\,. \label{l_n-def}
\end{eqnarray}
To find $b_{2B^*}^c$, we substitute expressions (\ref{h^0,1-def})
for $h_{\beta v}^0$, $h_{\beta B}^0$, $h_{\beta v}^1$, and
(\ref{tilde-m,t-expl-2}) for $\tilde{m}^q_{p0}$, into Eq.
(\ref{b_1,2B^*^c}). It yields:
\begin{eqnarray}
\hskip-5mm&&b_{2B^*}^c={1\over f_{B0}}\Big\{2(6y_1+x_4)x_2^2/D_0^2\nonumber\\
\hskip-5mm&&\hskip10mm-{[(3x_4-2y_1)+(6y_1+x_4)x_2/D_0]^2\over
(2y_1+7x_4)}\Big\}\,. \label{b_2B^*}
\end{eqnarray}

The total tracer solute enhancement factor is given by Eq.
(\ref{b_p^*-sum}) with $p^*=B^*$. As the mean frequency
$\omega_{\beta}$ and the activity factor $A^{ac}_{\beta}$ in
(\ref{D_beta-1}) coincide with those for chemical diffusion, terms
$b_{B^*}^{\omega}$ and $b_{B^*}^{ac}$ in (\ref{b_p^*-sum}) can be
found using Eqs.  (\ref{b_A,B^omega}) and (\ref{A_ac-PCA}) for a
binary alloy. Therefore, $b_{B^*}^{\omega}$ is equal to
$b_{B}^{\omega}$ in (\ref{b_A,B^omega}):
\begin{equation}
b_{B^*}^{\omega}=(4y_1+14x_4-18+b_{vB}+b_{BB})\, \label{b_B^*-omega}
\end{equation}
while expansion of Eq. (\ref{A_ac-PCA}) at low $c$ shows that
$b_{B^*}^{ac}$  is equal to the quantity $b_{BB}$ in (\ref{b_B}):
\begin{equation}
b_{B^*}^{ac}= b_{BB}=-12f_1^{BB}-6f_2^{BB}-\ldots\, \label{b_B^*-ac}
\end{equation}
Thus the tracer solute enhancement factor $b_{B^*}$ can be written
as
\begin{equation}
b_{B^*}=b_{B^*}^{\omega}+(b_{1B^*}^c +b_{2B^*}^c)+
b_{B^*}^{ac}\label{b_B^*-total}
\end{equation}
where various terms are given by Eqs.
(\ref{b_1B^*})\,-\,(\ref{b_B^*-ac}).

Experimental values of $b_{B^*}$ usually notably exceed unity:
$|b_{B^*})\gg 1$, similarly to $b_{A^*}$ values \cite{LeClaire-78}.
As in the case of $b_{A^*}$ discussed above, these large values can
imply that the main contribution to $b_{B^*}$ is made by the
frequency and activity terms in (\ref{b_B^*-total}),
$b_{B^*}^{\omega}$ and $b_{B^*}^{ac}$, as expressions
(\ref{b_B^*-omega})\, and \,(\ref{b_B^*-ac}) include large numerical
factors, while contributions of correlation terms $b_{1B^*}^c$ and
$b_{2B^*}^c$ in the total $b_{B^*}$  are usually less significant.

\section{ESTIMATES OF PARAMETERS OF FIVE-FREQUENCY MODEL FOR
REAL ALLOYS \label{estimates-of-x_n}}

Basic parameters of five-frequency model, $x_1$, $x_2$ and $x_4$ [or
frequency ratios $w_2/w_1$, $w_3/w_1$ and $w_4/w_0$ in
(\ref{w_n-e_i})] can be estimated from experimental data about the
ratio  of tracer diffusion coefficients,
$R_{D}^*=D_{B^*}^0/D_{A^*}^0$, the solute correlation factor
$f_{B0}$, and the vacancy flow factor $G=L_{AB,0}/L_{BB,0}$
\cite{Allnatt-93,Bocquet-96}. Then data about the tracer solvent
enhancement factor $b_{A^*}$ described by Eqs. (\ref{b_A^*c})-(\ref
{b_A^*-tot}) enable us to estimate the vacancy-solute interaction if
we suppose it to be short-ranged: $b_{vB}=-12f_1^{vB}$.

In table III we present estimates of $x_n$ and $v_1^{vB}$ for
several alloys for which data about $R^*_D$, $f_{B0}$, $G$ and
$b_{A^*}$ are available. For alloys AgZn, we present  $G$ estimated
from electromigration data in Ref. \cite{Doan-75}. For alloys AlZn,
we are not aware of data about $b_{A*}$, thus for $f_1^{vB}$ we give
its expression via this unknown $b_{A*}$. To calculate $R^*_D$ and
$f_{B0}$, we use the following relations:
\begin{equation}
R_{D}^*=x_2f_{B0}/f_0,\qquad G=L_{AB}^{c0}/(1-L_{BB}^{c0})
\label{R_D^*-G}
\end{equation}
where $f_0=0.7815$ is the exact correlation factor for tracer
self-diffusion \cite{Allnatt-93}, while $f_{B0}$ and $L_{pq}^{c0}$
are given by Eqs. (\ref{L-pq_0-nn})-(\ref{rho^0_i}) with $P_N$ and
$P_D$ taken from \cite{Manning-64}. For the tracer enhancement
factor $b_{A^*}$ we used Eqs. (\ref{b_A^*c})-(\ref{b_A^*-tot}).
Errors in estimates of interactions $v_1^{bV}$, $u_1^B$ and
$\Delta_A^B$ in Table III for Cu-based and Ag-based alloys
correspond to the variations of $|G|$ by $\pm 10\%$, while for AlZn
alloys these errors correspond to $\delta G=\pm 0.13$ given in
\cite{Hagenschulte-94}.

\begin{widetext}


\noindent TABLE III. Estimates of parameters of five-frequency model
for some FCC alloys from experimental data

\vspace{3mm}


\begin{tabular}{|cc|ccccc|cccc|ccc|}
\hline
&&\multicolumn{5}{c|}{}&\multicolumn{4}{c|}{}&\multicolumn{3}{c|}{}\\
&&\multicolumn{5}{c|}{Data used}&\multicolumn{4}{c|}{Parameters}
&\multicolumn{3}{c|}{Interactions, meV}\\
\hline &&&&&&Source&&&&&&&\\
Alloy&\hskip3mm $T$, K\hskip3mm&\hskip3mm$R_{D}^*$&\hskip3mm$f_{B0}$
\hskip3mm&\hskip3mm $G$\hskip3mm&\hskip3mm$b_{A^*}$\hskip3mm&of data
&\hskip3mm$x_1$ \hskip3mm&\hskip3mm$x_2$ \hskip3mm&\hskip3mm$x_4$
\hskip3mm&\hskip3mm$f_1^{vB}$\hskip3mm&\hskip3mm$v_1^{vB}$\hskip3mm&\hskip3mm$u_1^B$\hskip3mm
&\hskip3mm$\Delta_A^B$\hskip3mm\\
\hline &&&&&&&&&&&&&\\
CuZn&1168&3.56&0.47&-0.22&7.3&\cite{Allnatt-93,Bocquet-96}&2.2&5.9&1.25&0.04&\hskip2mm-4$\pm$
2&\hskip3mm23$\mp$2\hskip3mm&\hskip3mm-33$\mp$6\hskip3mm\\
&&&&&&&&&&&&&\\
CuCd&1076&10.2&0.22&-0.7&35&\cite{Allnatt-93,Bocquet-96}&5.1&36&2.5&0.1&\hskip2mm-9$\pm$23
&\hskip3mm85$\mp$12\hskip3mm&\hskip3mm18$\mp$31\hskip3mm\\
&&&&&&&&&&&&&\\
CuIn&1089&12&0.07&-0.57&43&\cite{Allnatt-93,Bocquet-96}&4.2&134&3.0&-0.12&\hskip2mm12$\pm$31
&\hskip3mm105$\mp$10\hskip3mm&\hskip3mm74$\mp$30\hskip3mm\\
&&&&&&&&&&&&&\\
CuSn&1089&14.1&0.15&-0.84&48&\cite{Allnatt-93,Bocquet-96}&6.6&73&3.4&0.57&\hskip2mm-42$\pm$27
&\hskip3mm115$\mp$18\hskip3mm&\hskip3mm52$\mp$48\hskip3mm\\
&&&&&&&&&&&&&\\
CuSn&1014&17&0.15&-1.06&40&\cite{Allnatt-93,Bocquet-96}&9.6&89&3.2&1.8&\hskip2mm-88$\pm$28
&\hskip3mm102$\mp$27\hskip3mm&\hskip3mm6$\mp$67\hskip3mm\\
\hline &&&&&&&&&&&&&\\
AgZn&1153&3.9&0.57&-0.39&12.7&\cite{Bocquet-96,Doan-75}&3.1&5.3&1.8&0.52&\hskip2mm-42$\pm$5
&\hskip3mm59$\mp$5
\hskip3mm&\hskip3mm6$\mp$14\hskip3mm\\
&&&&&&&&&&&&&\\
AlZn&829&3.5&0.5&-0.19&&\cite{Hagenschulte-94}&2.2&5.5&1.4&$(9.8-b_{A^*})/12$&&23$\pm$9
&-9$\pm$29\\
\hline
\end{tabular}
\vskip1mm
\end{widetext}

Before to discuss physical implications of results presented in
table IV we note that, according to usual ideas \cite{Allnatt-93},
an increase in excess of the valency  and atomic volume of impurity,
$Z_B$ and $\bar{v}_B$, with respect to those of host atoms,   $Z_A$
and $\bar{v}_A$, should lead, first, to the increase of the
vacancy-solute attraction as both the Coulomb and elastic
interactions become stronger and, second, to the increase of the
ratio $x_2=\omega_B^0/\omega_A^0$ as the activation energy
$E_{ac}^{Bv}$ in (\ref{gamma^pv}) should decrease as potential
minima for a $B$ atom in the host lattice at high
$\bar{v}_B/\bar{v}_A$ should become more shallow. Hence in the
sequence of alloys CuZn-CuCd-CuIn-CuSn we can expect, generally, an
increase of both the vacancy-solute attraction ($-v_1^{vB})$ and the
activation frequency ratio $x_2$, as well as the increase of this
ratio with lowering temperature $T$ .

The results for Cu-based alloys in table III, generally, agree with
these considerations (except CuIn alloys for which the experimental
value $f_{B0}$ seems to be abnormally low while errors in $v_1^{vB}$
are rather large). It can confirm that the five-frequency model
describes these alloys reasonably (even though it neglects many
physical effects, in particular, the long-ranged stress-induced
interactions which should be particularly important at high
$\bar{v}_B/\bar{v}_A$ \cite{VZh-12}). The kinetic interaction
$u_1^B$ defined by Eq. (\ref{u^alpha_il}) also increases in this
sequence of alloys which again seems to be natural. The saddle-point
interactions $\Delta_A^B$ in table III are usually weaker than
kinetic ones, and their changes with $Z_B/Z_A$ and
$\bar{v}_B/\bar{v}_A$ seem to be less pronounced.

Expressions (\ref{b_1B^*})\,-\,(\ref{b_B^*-total}) for the tracer
solute enhancement factor $b_{B^*}$ include also term $b_{BB}$
(\ref{b_B^*-ac}) which can be estimated from thermodynamic data,
see, e. g. \cite{Kulkarni-04}, and term $y_1$ in (\ref{x_n-def}).
Hence data about $b_{B^*}$ enable one to estimate the solute-solute
saddle-point interaction $\Delta_B^B$. It will be shown elsewhere
that the same six parameters: $x_1$, $x_2$, $x_4$, $b_{vB}$,
$b_{BB}$ and $y_1$, fully describe also the chemical (intrinsic)
diffusion enhancement factors, $b_A$ and $b_B$, for the
five-frequency model.

Unfortunately, we are not aware of reliable data about $b_{B^*}$ in
the ``fully-described'' alloys (such as those in table III) for
which data about  $R^*_D$, $f_{B0}$, $G$ and $b_{A^*}$ are
available. For example, Ref. \cite{Kulkarni-04} includes data about
both tracer and intrinsic diffusion coefficients, $D_{p^*}(c)$ and
$D_{p}(c)$, for CuZn at $T=1053$ K and AgCd at $T=873$ K. However,
values of $b_{A^*}$ in these data strongly differ from those
obtained by other authors at similar temperatures
\cite{Allnatt-93,Bocquet-96}, and the necessary relation
$D_{B^*}^0=D_{B}^0$ is notably violated. The reliable data about
$D_{B^*}$, $D_{A}$ and $D_{B}$ in the ``fully-described'' FCC alloys
will allow to estimate all six microscopic parameters of the theory.

\section{CONCLUSIONS\label{Conclusions}}

Let us summarize the main results of this work. We present the new
formulation of the  master equation approach to the theory of
diffusion in substitution alloys using the five-frequency model of
FCC alloys as an example. Unlike the earlier version of this
approach suggested by Nastar et al.
\cite{Nastar-00,Nastar-05,Barbe-06}, our formulation gives the
explicit form for all equations of the theory and uses the
well-elaborated methods of statistical physics to approximately
solve these equations. The approach developed is used to calculate
the enhancement factors for tracer solvent and tracer solute
diffusion in dilute FCC alloys. We show that some significant
contribution to the tracer solvent enhancement factor related to the
vacancy-solute interaction was missed in the previous treatments of
this problem. It implies that existing estimates of parameters of
five-frequency model for the most of real alloys should be revised.
For several FCC alloys for which necessary experimental data are
available, we estimate these parameters, including the
vacancy-solute interaction. The results obtained seem to show that
the five-frequency model for these alloys is adequate. We also
discuss the experiments needed to fully describe both tracer and
chemical diffusion in FCC alloys in the framework of the
five-frequency model.

\section*{ACKNOWLEDGEMENTS}

The work was supported by the Russian Fund of Basic Research (grant
No. 12-02-00093); by the fund for support of leading scientific
schools of Russia  (grant No. NS-215.2012.2); and by the program of
Russian university scientific potential development (grant  No.
2.1.1/4540).

\end{document}